\documentclass[twocolumn]{aastex61}
\usepackage{hyperref}

\shorttitle{Nucleosynthesis in core-collapse supernovae}

\shortauthors{Wanajo et al.}

\begin{document}

\title{Nucleosynthesis in the innermost ejecta of neutrino-driven supernova
explosions in two dimensions\\
}

\newcommand{\X}[1]{ #1}
\newcommand{\R}[1]{ #1}

\author{Shinya Wanajo}
\affil{Department of Engineering and Applied Sciences,
Sophia University, Chiyoda-ku, Tokyo 102-8554, Japan;
shinya.wanajo@sophia.ac.jp}
\affil{iTHES Research Group, RIKEN, Wako, Saitama 351-0198, Japan}

\author{Bernhard M\"uller}
\affil{Astrophysics Research Centre, School
  of Mathematics and Physics, Queens University Belfast, Belfast, BT7
  1NN, United Kingdom}
\affil{Monash Centre for Astrophysics, School of Physics and
  Astronomy, Monash University, VIC 3800, Australia}

\author{Hans-Thomas Janka}
\affil{Max-Planck-Institut f\"ur Astrophysik,
        Karl-Schwarzschild-Str. 1, D-85748 Garching, Germany}

\author{Alexander Heger}
\affil{Monash Centre for Astrophysics, School of Physics and
  Astronomy, Monash University, VIC 3800, Australia}
\affil{University of Minnesota, School of Physics and Astronomy, Minneapolis, MN 55455,
USA}
\affil{Shanghai Jiao-Tong University, Department of Physics and Astronomy, Shanghai
200240, P.~R.~China}

\begin{abstract}

We examine the nucleosynthesis in the innermost, neutrino-processed
ejecta (a few $10^{-3}\,M_\odot$) of self-consistent, two-dimensional
explosion models of core-collapse supernovae for six progenitor stars with
different initial masses. Three models have initial masses near the
low-mass end of the supernova range, $8.8\,M_\odot$ (e8.8;
electron-capture supernova), $9.6\,M_\odot$ (z9.6), and $8.1\,M_\odot$
(u8.1), with initial metallicities of 1, 0, and $10^{-4}$ times the
solar metallicity, respectively. The other three are solar-metallicity
models with initial masses of $11.2\,M_\odot$ (s11),
$15\,M_\odot$ (s15), and $27\,M_\odot$ (s27). 
The low-mass models e8.8, z9.6, and u8.1 exhibit
high production factors (nucleosynthetic abundances relative to the
solar ones) of 100--200 for light trans-iron elements from Zn to Zr. This
is associated with appreciable ejection of neutron-rich matter in these
models. Remarkably, the nucleosynthetic outcomes for progenitors
e8.8 and z9.6 are
almost identical, including interesting productions of $^{48}$Ca and
$^{60}$Fe, irrespective of their quite different (O-Ne-Mg and Fe)
cores prior to collapse. In the more massive models s11, s15,
and s27, several proton-rich isotopes of light trans-iron elements,
including the $p$-isotope $^{92}$Mo (for s27) are made, up to production
factors of $\sim$30. Both electron-capture and core-collapse supernovae
near the low-mass end can therefore be dominant contributors to the
Galactic inventory of light trans-iron elements from Zn to Zr and
probably $^{48}$Ca and live $^{60}$Fe. The innermost ejecta of more
massive supernovae may have only sub-dominant contributions to the
chemical enrichment of the Galaxy except for $^{92}$Mo.




\end{abstract}


\section{Introduction}

Core-collapse supernovae (CCSNe), the deaths of stars with initial
masses heavier than about $8 M_\odot$, have long been suggested to be
important astrophysical sources of trans-iron species as well as of
intermediate-mass and iron-group elements. Traditionally, studies of
CCSN nucleosynthesis were based on artificial  one-dimensional (1D) explosion
models with free parameters such as the mass-cut (location of
ejecta-remnant interface), the electron fraction ($Y_\mathrm{e}$; number
of protons per nucleon) of the ejecta, and the explosion energy
\citep[e.g.,][]{Woosley1995, Thielemann1996, Rauscher2002,
Limongi2006, Tominaga2007, Heger2010, Nomoto2013}. Nucleosynthetic
abundances made near the mass-cut in the innermost ejecta  are,
however, highly dependent on such free parameters. 

The approach of replacing
  the complex multi-dimensional dynamics of a neutrino-driven supernova (SN)
  with a piston or thermal bomb as an artificial explosion engine
  is likely to be problematic when it comes to studying the innermost ejecta. 
  \X{These parameterized models face two
  serious problems in modelling the innermost
  ejecta: Most importantly, they
  fail to reproduce the thermodynamic history
  of the neutrino-driven ejecta, i.e.,
  of material that makes its way deep into
  the gain region, is then heated by neutrinos
  and ejected in buoyant high-entropy bubbles.
  In these ejecta, any memory of the initial
  composition is lost as 
  charged-current neutrino interactions
  \emph{completely reset} the electron fraction
  $Y_\mathrm{e}$ and entropy $s$.
  The final $Y_\mathrm{e}$ and $s$ in the ejecta
  are determined by the interplay of
  the neutrino reactions and
  the expansion of the ejecta (see
  \citealt{Pruet2005,Pruet2006,Froehlich2006,Wanajo2011a} and in particular Section~2.2.1 of \citealt{Mueller2016}).
  It is therefore
  impossible to reproduce this ejecta
  component if one assumes  that the only effect of the supernova
  engine on the ejecta is to provide
  shock heating as in simpler
  models of the engine. Moreover,
  even for the explosive burning of
  material that is immediately ejected
  after being shocked, the simple
  engine models are problematic for
  material close to the ``mass cut'':
  For ejecta that are shock-heated
  while the supernova engine
  is still feeding energy into
  the incipient explosion
  (which can take seconds;
  \citealp{Mueller2015,Bruenn2016,Mueller2017})},
  the shock heating is sensitive
  to the nature of the engine,
  and it is therefore not sufficient
  to merely tune an artificial engine
  to produce plausible \emph{final}
  explosion energies and nickel masses to
  capture the nucleosynthesis in this ejecta
  component.
This makes it difficult to predict
nucleosynthesis of some iron-group and trans-iron (Zn and heavier)
elements based on such parameterized models
\X{in these two ejecta components, which we
term the ``innermost ejecta''.
It is therefore imperative to complement
extant studies of the detailed explosive
and hydrostatic nucleosynthesis
with nucleosynthesis calculations
based on more consistent models
of the supernova engine.
}

\begin{deluxetable*}{cccccccccc}
\tabletypesize{\scriptsize}
\tablecaption{Parameters of Supernova Explosion Models}
\tablewidth{0pt}
\tablehead{
  \colhead{Model} &
  \colhead{$M_\mathrm{prog}$\tablenotemark{a}} &
  \colhead{$Z_\mathrm{prog}$\tablenotemark{b}} &
  \colhead{\X{$M_\mathrm{CO}$\tablenotemark{c}}} &
  \colhead{\X{$M_\mathrm{He}$\tablenotemark{d}}} &
  \colhead{$\Delta \theta$\tablenotemark{e}} &
  \colhead{$t_\mathrm{expl}$\tablenotemark{f}} &
  \colhead{$t_\mathrm{fin}$\tablenotemark{g}} &
  \colhead{\X{$m_\mathrm{fin}$\tablenotemark{h}}} &
  \colhead{$E_\mathrm{diag}$\tablenotemark{i}} \\
  \colhead{} &
  \colhead{($M_\odot$)} &
  \colhead{($Z_\odot$)} &
  \colhead{\X{($M_\odot$)}} &
  \colhead{\X{($M_\odot$)}} &
  \colhead{deg} &
  \colhead{(ms)} &
  \colhead{(ms)} &
  \colhead{($M_\odot$)} &
  \colhead{($10^{50} \, \mathrm{erg}$)} }
  \startdata
  e8.8  & 8.8 & 1                  & \X{1.377} & \X{1.377} & 1.4 & 92   & 362  & \X{$\mathord{>}1.377$\tablenotemark{j} }
  & 0.9 \\
  z9.6  & 9.6 & 0                  & \X{1.371} & \X{1.69} & 1.4 & 125  & 1420 & \X{$\mathord{>}1.373$}\tablenotemark{k}  & 0.6\\
  u8.1  & 8.1 & $10^{-4}$          & \X{1.385} & \X{1.92} & 1.4 & 177  & 335  & \X{1.373} & 0.4 \\
s11\tablenotemark{l}   & 11.2& 1   & \X{1.89}  & \X{2.84} & 2.8 & 213  & 922  & \X{1.39} & 0.4 \\
s15\tablenotemark{m} & 15  & 1     & \X{2.5}   & \X{4.5 } & 2.8 & 569  & 779  & \X{1.45} & 1.3 \\
s27\tablenotemark{n}     & 27  & 1 & \X{7.2}   & \X{9.2 } & 1.4 & 209  & 790  & \X{1.74} & 1.9 \\
\enddata
\tablenotetext{a}{Progenitor mass at the zero-age main sequence.}
\tablenotetext{b}{Metallicity at the zero-age main sequence.}
\tablenotetext{c}{\X{CO core mass at collapse.}}
\tablenotetext{d}{\X{He core mass at collapse.}}
\tablenotetext{e}{Angular resolution.}
\tablenotetext{f}{Post-bounce time of explosion, defined as the point in time when the average shock radius $\langle r_\mathrm{sh} \rangle$ reaches $400 \ \mathrm{km}$.}
\tablenotetext{g}{Final post-bounce time reached in simulation.}
\tablenotetext{h}{Mass coordinate corresponding to the average shock radius 
at the end of the simulation.}
\tablenotetext{i}{Diagnostic explosion energy at the end of simulation.}
\tablenotetext{j}{\X{Note
that an artificial envelope with
$\rho \propto r^{-3}$ was used in
\citet{Wanajo2011a} outside the He core. It is therefore
not appropriate to
give a more precise mass coordinate for
final position of the shock in the hydrogen shell.}}
\tablenotetext{k}{By the end of the simulation,
the shock has crossed the outer grid boundary
at a mass coordinate of $1.373 M_\odot$ in model z9.6.}
\tablenotetext{l}{Same as model s11.2 in \citet{Woosley2002} and \citet{Mueller2012a}.}
\tablenotetext{m}{Same as model s15s7b2 in \citet{Woosley1995} and \citet{Mueller2012a}.}
\tablenotetext{n}{Same as model s27.0 in \citet{Woosley2002} and \citet{Mueller2012b}.}
\label{tab:models}
\end{deluxetable*}

\begin{deluxetable*}{ccccccccc}
\tabletypesize{\scriptsize}
\tablecaption{Properties of Innermost SN Ejecta}
\tablewidth{0pt}
\tablehead{
\colhead{Model} &
\colhead{Type} &
\colhead{$M_\mathrm{PNS}$\tablenotemark{c}} &
\colhead{$M_\mathrm{ej}$\tablenotemark{d}} &
\colhead{$M_\mathrm{ej, n}$\tablenotemark{e}} &
\colhead{$Y_\mathrm{e, min}$\tablenotemark{f}} &
\colhead{$Y_\mathrm{e, max}$\tablenotemark{g}} &
\colhead{$S_\mathrm{min}$\tablenotemark{h}} &
\colhead{$S_\mathrm{max}$\tablenotemark{i}} \\
\colhead{} &
\colhead{} &
\colhead{($M_\odot$)} &
\colhead{($10^{-3} M_\odot$)} &
\colhead{($10^{-3} M_\odot$)} &
\colhead{} &
\colhead{} &
\colhead{($k_\mathrm{B}\, \mathrm{nuc}^{-1}$)} &
\colhead{($k_\mathrm{B}\, \mathrm{nuc}^{-1}$)}
}
\startdata
e8.8 & ECSN & 1.36  & 11.4 & 5.83  & 0.398 & 0.555 & 9.80 & 383\\ 
z9.6 & CCSN & 1.36  & 12.4 & 4.94  & 0.373 & 0.603 & 12.6 & 27.8\\ 
u8.1 & CCSN & 1.36  & 7.69 & 3.24  & 0.399 & 0.612 & 9.83 & 29.9\\ 
s11& CCSN   & 1.36  & 14.1 & 0.133 & 0.474 & 0.551 & 6.64 & 34.7\\ 
s15  & CCSN & 1.58  & 15.9 & 0.592 & 0.464 & 0.598 & 6.78 & 36.7\\
s27  & CCSN & 1.65  & 27.3 & 0.759 & 0.387 & 0.601 & 5.19 & 44.0\\
\enddata
\tablenotetext{c}{Baryonic mass of the proto-neutron star at the end of simulation.}
\tablenotetext{d}{Total mass in the innermost ejecta.}
\tablenotetext{e}{Ejecta mass with $Y_\mathrm{e} < 0.4975$.}
\tablenotetext{f}{Minimal $Y_\mathrm{e}$ evaluated at $T_9 = 10$ (see text).}
\tablenotetext{g}{Maximal $Y_\mathrm{e}$ evaluated at $T_9 = 10$ (see text).}
\tablenotetext{h}{Minimal asymptotic entropy (at the end of simulation).}
\tablenotetext{i}{Maximal asymptotic entropy (at the end of simulation).}
\label{tab:properties}
\end{deluxetable*}

Recent work by \citet{Sukhbold2016} has advanced the 1D approach by
adopting parameterized neutrino-powered explosion models, \X{which is a first step towards
addressing these problems}. In their
models the boundary conditions (e.g., proto-neutron star contraction
and core luminosity) are set deep inside the proto-neutron star  and the
physical conditions of the innermost ejecta are obtained as a result of
hydrodynamical computations \citep[see also][for comparable approaches
    beyond piston and thermal bomb models in 1D]{Froehlich2006, Perego2015}.
Similar to previous works,
they 
found production of elements from B to Cu in reasonable agreement with
the solar ratio (except for some elements that had other contributors
such as low-mass stars and SNe~Ia).
They reported, however, a severe deficiency of light trans-iron
elements from Zn to Zr, which had been explained in part by the weak
$s$-process in previous studies \citep[e.g.,][]{Woosley2007}. In
addition, astrophysical sources of $^{48}$Ca (neutron-rich isotope of
Ca), $^{64}$Zn (main isotope of Zn), and $^{92}$Mo ($p$-isotope) still
remain unresolved, for which a rare class of SNe~Ia \citep{Meyer1996,
Woosley1997}, hypernovae \citep{Umeda2002, Tominaga2007} and
neutron-rich nuclear equilibrium \citep{Hoffman1996, Wanajo2006},
respectively, have been proposed as possible explanations. Note, however, that \citet{Sukhbold2016} did not include neutrino-heated (but just shock-heated) ejecta in their analysis.

One of the fundamental problems in such 1D models is  obviously the
limitation of dimensionality.
Among other things, they do not account for the spatial variations
  in electron fraction $Y_\mathrm{e}$, entropy $S$, and ejection velocity that
is seen in multi-dimensional CCSN models. 
Two-dimensional (2D) simulations  with sophisticated neutrino
transport
therefore provide a much better basis for studying the 
nucleosynthesis in
the innermost CCSN ejecta. \citet{Pruet2005, Pruet2006} used the
hydrodynamical trajectories of a $15 M_\odot$ CCSN \citep{Buras2006}
for nucleosynthesis and showed that interesting amounts of Sc, Zn, and
light $p$-nuclei were formed in the innermost proton-rich ejecta
\citep[see also][]{Fujimoto2011}, while the contribution of neutron-rich ejecta appeared to be subdominant \citep{Hoffman2007}. 
This was an important step beyond
previous CCSN nucleosynthesis studies, but the explosion, though 
neutrino-driven, was still
induced artificially like the piston-induced or thermal-bomb explosions
of the previous 1D approaches, i.e., the 2D explosion was not 
obtained in a fully
self-consistent manner. 
Moreover, based on the superior treatment of the neutrino transport in
our present models we can re-investigate the question of nucleosynthesis in the
neutron-rich ejecta, in which light trans-iron species could be
abundantly produced \citep[e.g.,][]{Hoffman1996}.


To our knowledge, the only extant study of nucleosynthesis based on a
self-consistent multi-dimensional explosion model is the one  by \citet{Wanajo2011a,
Wanajo2013a, Wanajo2013b}, which is based on a  2D simulation of an
$8.8 M_\odot$ electron-capture SN \citep[ECSN; a sub-class of CCSNe
arising from collapsing O-Ne-Mg cores,][]{Nomoto1987, Janka2008}. They found appreciable production of trans-iron elements
from Zn to Zr, $^{48}$Ca, and $^{60}$Fe (radioactive nuclei) in the
innermost neutron-rich ejecta, a very different result from
nucleosynthesis studies
  \citep{Hoffman2008, Wanajo2009}
  based on the corresponding 1D models \citep{Kitaura2006,Janka2008}. This demonstrates the importance
of self-consistent multi-D models for
  reliable nucleosynthesis predictions.

2D explosion models only provide a first glimpse at the
  role of multi-D effects in supernova nucleosynthesis, however. 
  Recent
  three-dimensional (3D) core-collapse simulations have shown
  qualitative and quantitative differences to 2D \citep[see, e.g.,][for a recent review]{Janka2016} concerning
  shock revival (with a trend towards delayed or missing explosions)
  and also concerning the multi-D flow dynamics and energetics during the first phase of the explosion
\citep{Melson2015, Mueller2015}. 
Thus,  nucleosynthesis studies based on 3D
models will eventually be needed, but are, of course, computationally
much more  demanding.

This paper aims at extending the 2D studies of \citet{Wanajo2011a,
Wanajo2013a, Wanajo2013b} to examine the nucleosynthesis in the innermost
ejecta of CCSNe arising from iron-core progenitors.
This will be an
important step toward future nucleosynthesis studies from
  self-consistent 3D explosion models. In addition to the ECSN model
studied by \citet{Wanajo2011a,Wanajo2013a,Wanajo2013b}, we consider five  CCSN models with 
zero-age main sequence  progenitor masses of $9.6\, M_\odot$ (initial metallicity of $Z=0\,
Z_\odot$), $8.1\, M_\odot$ ($10^{-4}\, Z_\odot$), and $11.2\,
M_\odot$, $15\, M_\odot$, and $27 M_\odot$ ($1\, Z_\odot$)
(\S~\ref{sec:ccsnmodels}).
By including the $8.1 M_\odot$ and $9.6 M_\odot$ progenitors side-by-side with an
  ECSN model, our study retains
  a strong focus on the low-mass end of the progenitor spectrum.
  With the uncertainties surrounding the ECSN channel \citep{Poelarends2008,Jones2013,Jones2014,Doherty2015,Jones2016},
  one particular
  question that we seek to answer is whether ECSN-like nucleosynthesis
  can also be obtained for slightly different progenitor channels close
  to the iron core formation limit.

All these SN explosions were
obtained self-consistently, i.e., with no free parameters, in 2D
axisymmetric simulations adopting an elaborate neutrino transport
scheme \citep{Mueller2012a, Mueller2012b, Janka2012}. Nucleosynthetic
abundances are calculated by applying an up-to-date reaction network
code to hydrodynamic trajectories in a post-processing step
(\S~3). Some details of the nucleosynthesis mechanisms operating in the
innermost ejecta are described by taking the result of the metal-free
$9.6\, M_\odot$ star as representative of our models. The
mass-integrated nucleosynthetic yields are compared to the solar
abundances to test if the innermost ejecta of these SNe can be major
sources of light trans-iron elements and some other species in the
Galaxy (\S~4). 
Our conclusions follow (\S~5).

\section{SN models}\label{sec:ccsnmodels}

\subsection{Numerical methods and progenitor models }

All the SN hydrodynamical trajectories have been computed
from 2D general relativistic simulations with
  energy-dependent ray-by-ray-plus neutrino transport based
    on a variable Eddington factor technique as implemented
    in the supernova code \textsc{Vertex} \citep{Rampp2002,Buras2006,Mueller2010}.
   Except for the pseudo-relativistic
    ECSN model of \citet{Wanajo2011a}, general relativity is treated
    in the extended conformal flatness approximation \citep{Cordero2009}.
    A modern set of neutrino interaction rates
    (the ``full rates'' set of \citealt{Mueller2012a}) has been used for the simulations.   
The explosions were obtained
self-consistently and thus the models contained no free parameters. 
  

The initial pre-SN models were adopted from
\citet[][$8.8\, M_\odot$ star with an O-Ne-Mg core with solar
  metallicity]{Nomoto1987},
A. Heger (unpublished\footnote{Extension of \citet{Heger2010}.}, $9.6\, M_\odot$ and
$8.1\, M_\odot$ models\footnote{Both the $9.6\,M_\odot$ of zero metallicity and the $8.1\,
  M_\odot$ model of $10^{-4}\, Z_\odot$ were at the
  low-mass ends of CCSN progenitors for their respective metallicities in the
  KEPLER calculations.  See also \citet{Ibeling2013} for general metallicity
  dependence of the CCSN lower mass limit.} with metallicities $Z=0$ and $Z=10^{-4}\, Z_\odot$,
respectively), \citet[][$11.2\, M_\odot$ solar metallicity model,
s11.2 in their paper]{Woosley2002}, \citet[][$15\, M_\odot$ solar
metallicity model, s15s7b2 in their paper]{Woosley1995},
\citet[][$27\, M_\odot$ solar metallicity model, s27.0 in their
paper]{Woosley2002}. The corresponding explosion models are labeled as
e8.8, z9.6, u8.1, s11, s15, and s27 hereafter.
Relevant model parameters are summarized in
  Table~\ref{tab:models}.
  For more details on the individual
  explosion models, we refer the reader to the original publications
  on e8.8 \citep{Wanajo2011a}, z9.6 \citep{Janka2012}, u8.1 \citep{Mueller2012b},
  s11, s15s7b2 \citep{Mueller2012a}, and s27 \citep{Mueller2012b}.

  Note that although we include two models of subsolar metallicity,
    the limited number of models does not permit us to discuss the
    dependency on metallicity in this paper. 
    

\begin{figure*}
\epsscale{0.5}
\plotone{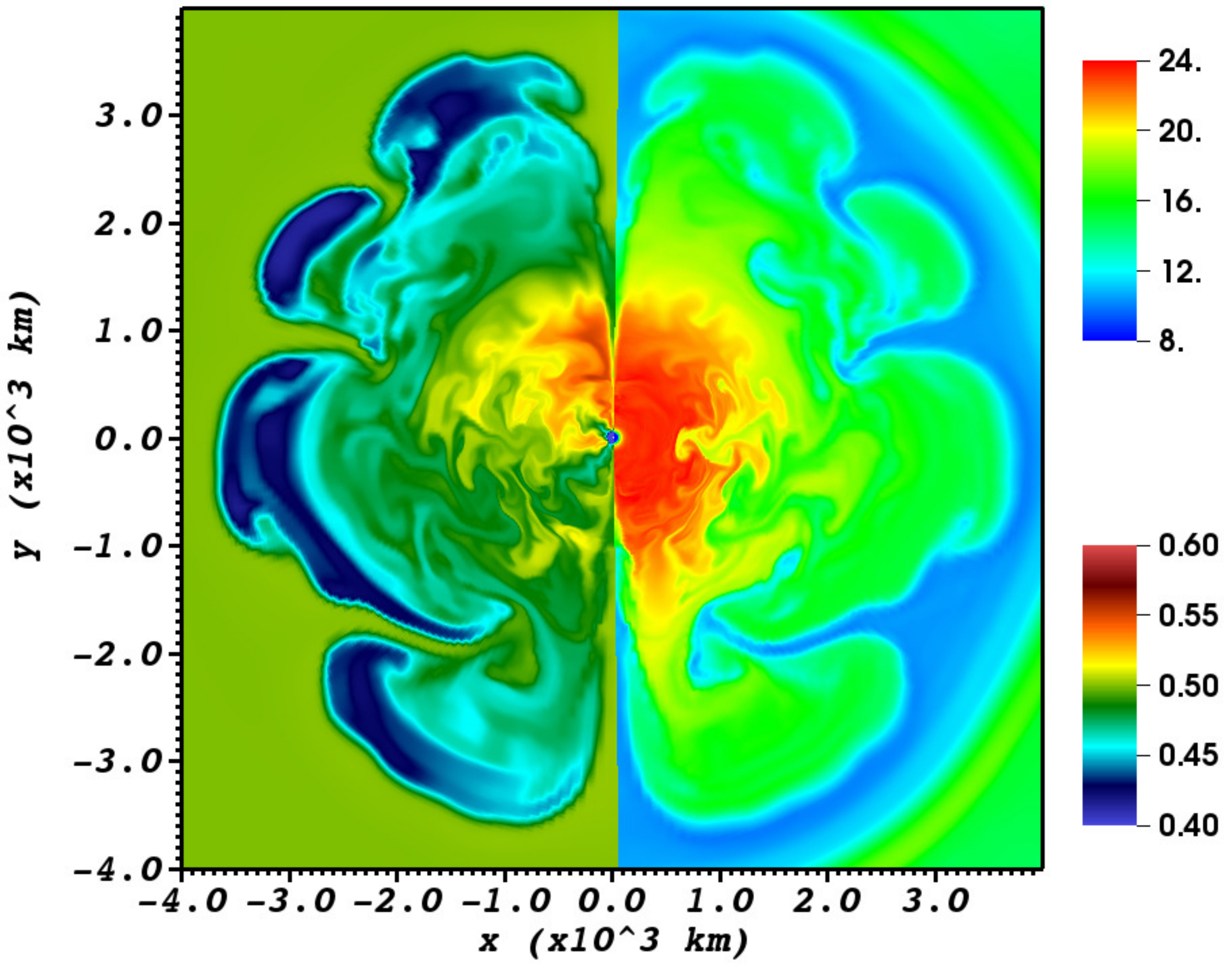}
\plotone{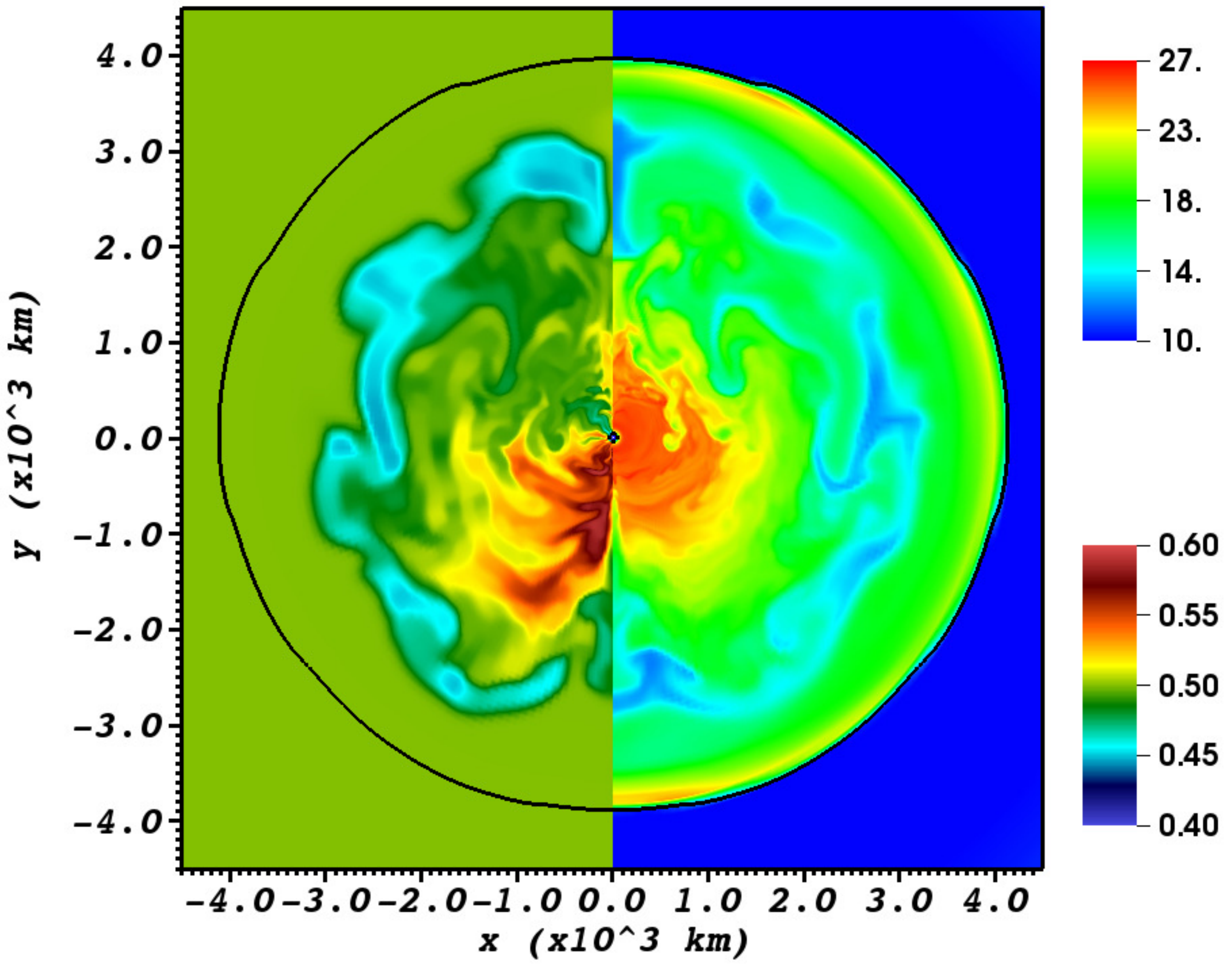}
\plotone{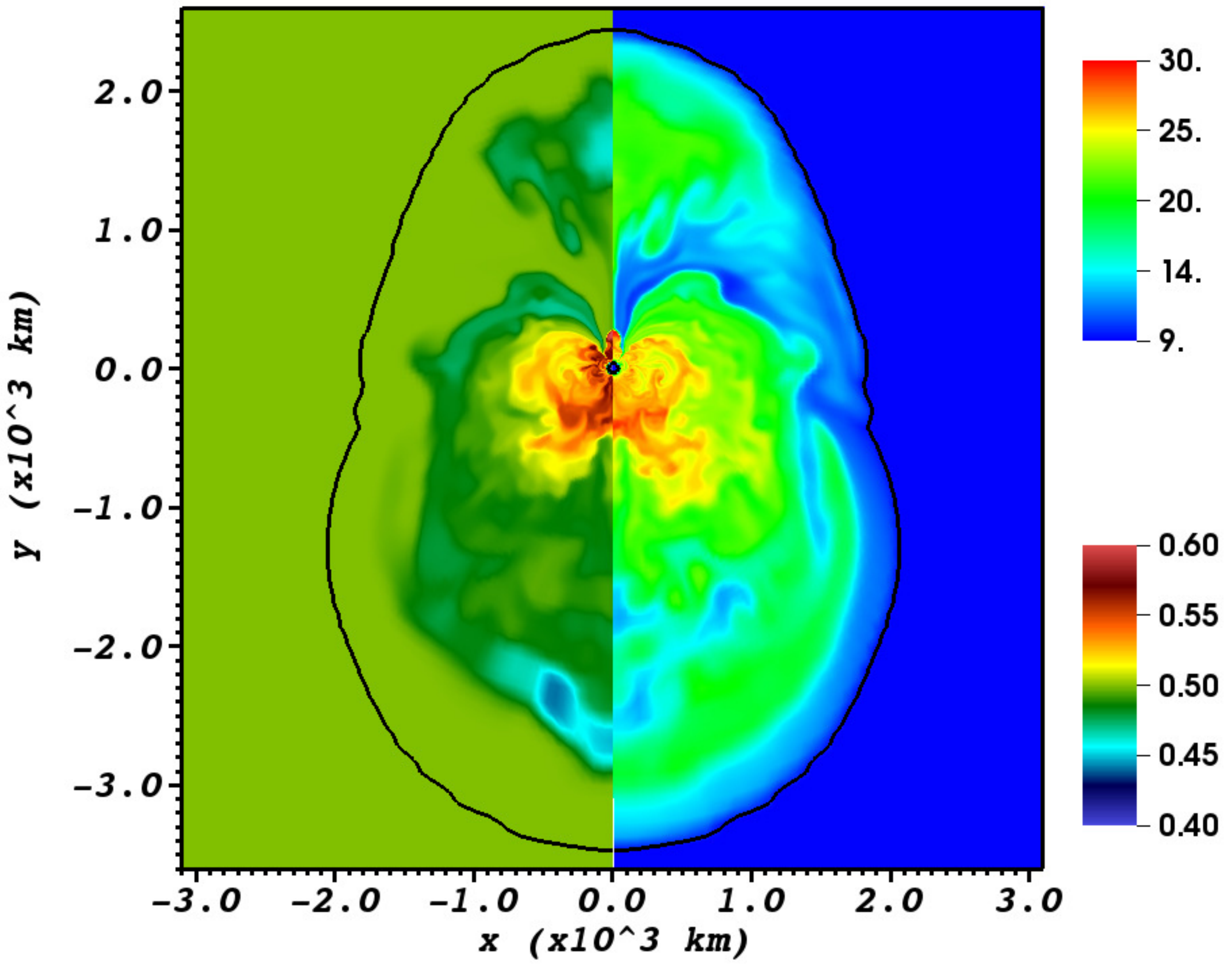}
\plotone{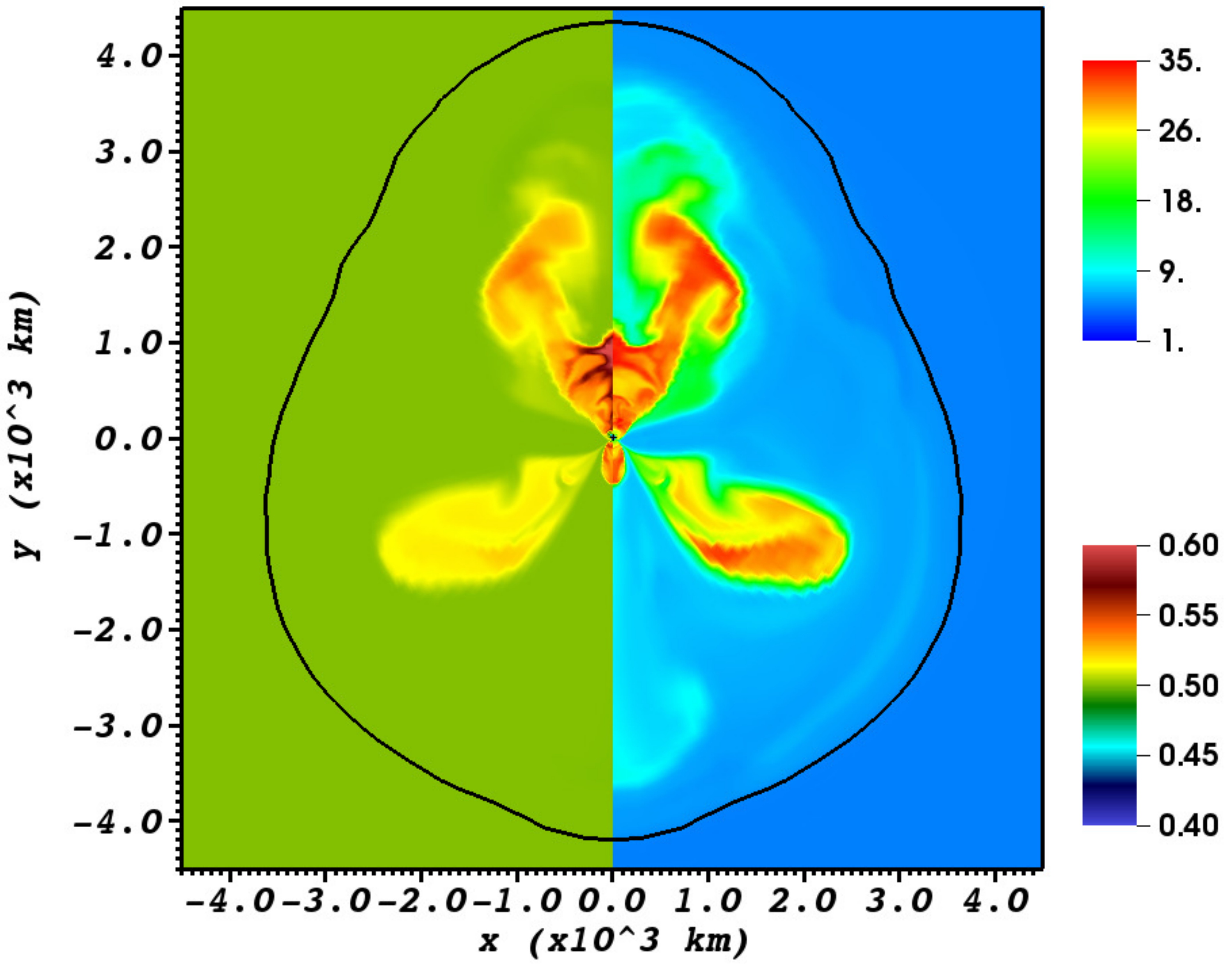}
\plotone{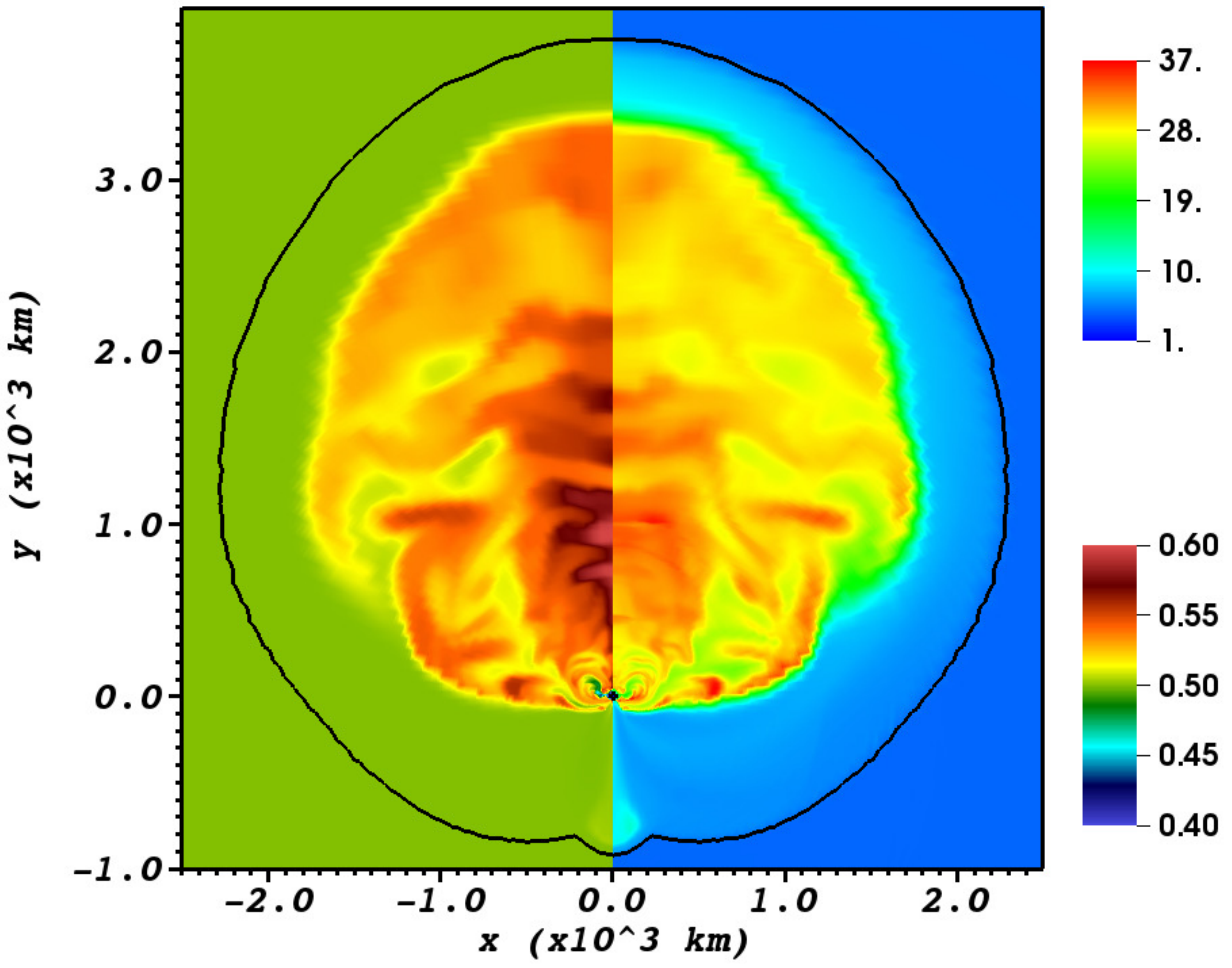}
\plotone{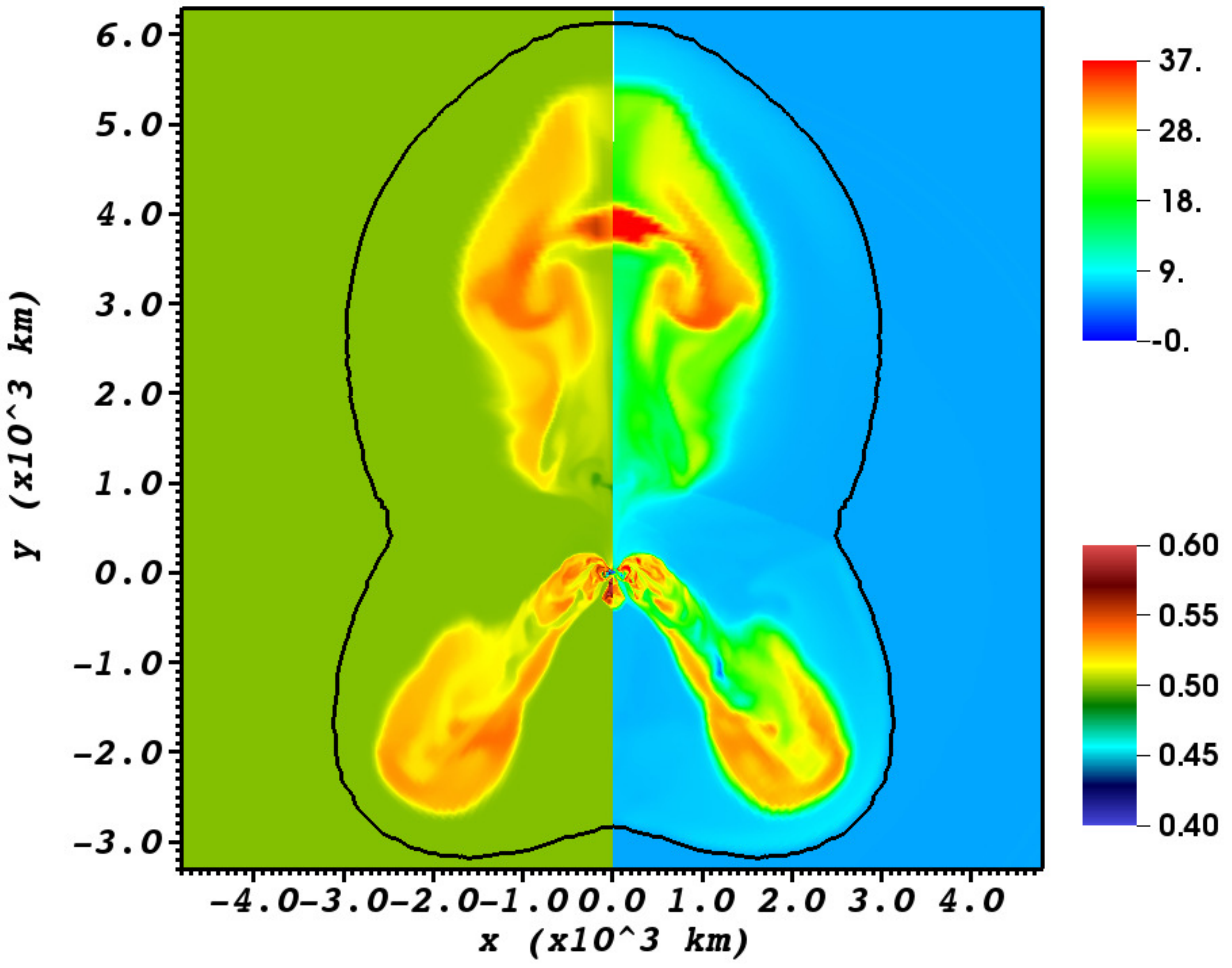}

\caption{Late-time snapshots of models e8.8 (left-top, $266\,
\mathrm{ms}$ after bounce), z9.6 (right-top, $317\, \mathrm{ms}$), u8.1
(left-middle, $315\, \mathrm{ms}$), s11 (right-middle, $922\,
\mathrm{ms}$), s15 (left-bottom, $776\, \mathrm{ms}$), and s27
(right-bottom, $790\, \mathrm{ms}$). Each panel shows the distribution
of electron fraction, $Y_\mathrm{e}$ (left halves of panels; color-bar
at right-bottom), and entropy per nucleon, $S$, in units of
$k_\mathrm{B}/\mathrm{nucleon}$ (right; color-bar at right-top) at a
time when the high-entropy plumes of neutrino-heated matter have reached
a radius of roughly $3000 \, \mathrm{km}$.  The black line in each panel
indicates the shock front. The vertical and horizontal axes show the
distance from the center.  Note that the shock has already propagated
well beyond $4000 \, \mathrm{km}$ in e8.8 at this stage and is therefore
no longer visible in the plot. } \label{fig:snimage}
\end{figure*}

\subsection{Explosion dynamics }
Snapshots of the electron fraction, $Y_\mathrm{e}$, and
the entropy per nucleon, $S$, in these simulations at late times are shown in
Figure~\ref{fig:snimage}.
One can see roughly spherical structures for e8.8 and z9.6 (top panels) and more
  strongly asymmetric features with a dipolar or quadrupolar geometry for the other
models (middle and bottom panels).
The different explosion dynamics  reflects the
core structures of pre-SN stars with steeper to shallower density
gradients in the order of e8.8, z9.6, u8.1, s11, s15, and s27
\citep[see Fig.~8 in][]{Janka2012}. The ECSN progenitor for e8.8 is a
super-asymptotic giant branch (SAGB) star with an O-Ne-Mg core
surrounded by a very dilute H-He envelope, while those of CCSNe are
iron cores embedded by dense oxygen-silicon shells. For e8.8, therefore,
the explosion sets in very early at a post-bounce  time of
$t_\mathrm{pb} \sim 80$~ms before vigorous convection can develop. Overturn driven
by the Rayleigh-Taylor instability only occurs when the explosion is underway, but
the plumes do not have sufficient time to merge into large structures so that
no global asymmetry emerges. For the CCSN cases (except for z9.6), by contrast, the
explosions gradually start after multi-dimensional effects, i.e.,
convection or the standing-accretion-shock instability \citep[SASI,][]{Blondin2003},
have
reached the non-linear regime (the onset of the explosion occurs at $t_\mathrm{pb} \sim 500$~ms for s15 and
$t_\mathrm{pb} \sim 150\ldots 200$~ms for the others,
Table~\ref{tab:models} and
Fig.~14
in \citet{Janka2012}). Moreover, due to the presence of a relatively dense and massive
oxygen shell, shock expansion is slow enough for a dominant unipolar or bipolar
asymmetry to emerge after shock revival. 

Models z9.6 and u8.1 stand apart from the more massive CCSN models s11, s15, and s27,
  since stars close to the iron core formation limit exhibit evolutionary and
  structural similarities to ECSN progenitors \citep{Jones2013,Jones2014,Woosley2015}.
  Specifically, these progenitors have very thin O and C shells between the core
  and the low-density He and H envelope.
 Because of these peculiarities, model z9.6 (and to some degree u8.1) is a case on the borderline
to ECSN-like explosion behavior \X{\citep{Mueller2016}}. Its progenitor  exhibits the steepest
core-density gradient near the core-envelope interface among the CCSN
cases, resulting in \X{a similarly rapid expansion of the neutrino-heated ejecta
and hence
in $Y_\mathrm{e}$ and $s$ structures}
 rather similar to those of e8.8. The progenitor of u8.1 has a
slightly shallower core-density gradient than that of z9.6 so that
the propagation of the shock is slightly slower than for z9.6, but
still faster than for typical iron-core progenitors. Since the width of
the ECSN channel is subject to considerable uncertainties
\citep[see][and references therein]{Poelarends2008,Jones2013,Jones2014,Doherty2015,Jones2016},
CCSNe from this mass range are particularly interesting
as a possible alternative source for ``ECSN-like'' nucleosynthesis
\X{(i.e., the characteristic nucleosynthesis
in neutron-rich neutrino-heated
ejecta with moderate entropy
as discussed in \citealt{Wanajo2011a,Wanajo2013a,Wanajo2013b})}. 

\begin{figure*}
\epsscale{1.0}
\plotone{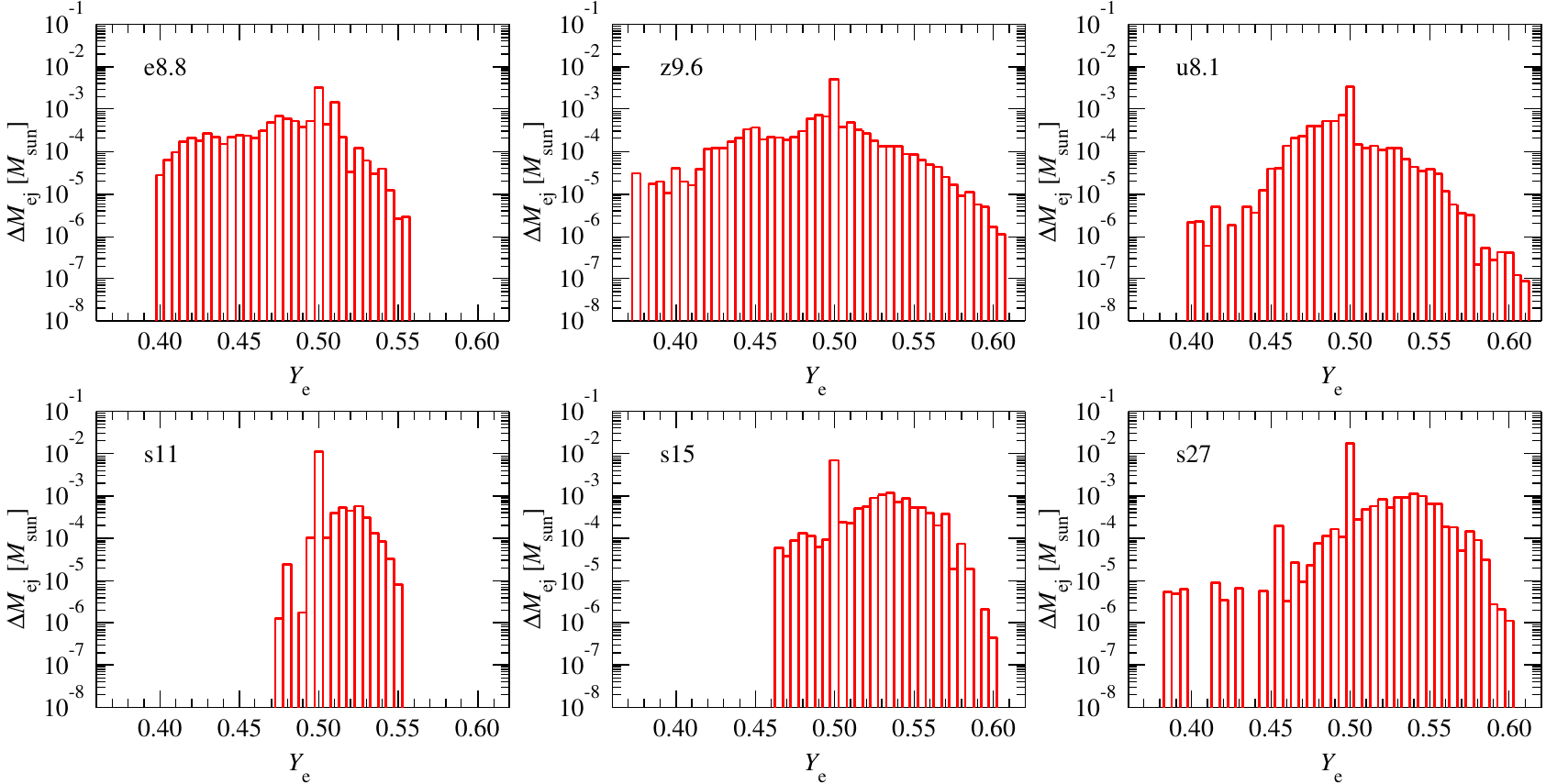}

\caption{$Y_\mathrm{e}$ histograms for the SN ejecta when the
 temperatures have decreased to $T_9 = 10$. The ejecta masses ($\Delta
 M_\mathrm{ej}$) are shown as functions of $Y_\mathrm{e}$ with a bin size 
 of $\Delta Y_\mathrm{e} = 0.005$.}

\label{fig:yehist}
\end{figure*}

As in \citet{Wanajo2011a} and similar to
 \citet{Wongwathanarat2016}, we compute the trajectories for our
  nucleosynthesis calculations from the 2D data files (with a time
  spacing of $0.25 \, \mathrm{ms}$) instead of co-evolving tracer
  particles during the simulation. In order not to follow the tracers
  during multiple convective overturns with the risk of accumulating
  discretization errors, we integrate the tracer trajectories
  \emph{backwards} in time from an appropriate point during the
  simulation (i.e., starting either from the end of the simulation, or
  from a time when the bulk of the ejecta have cooled sufficiently as
  in the case of z9.6).  This procedure greatly reduces the number of
  required trajectories to adequately sample the ejecta: Selecting the
  final locations of the tracer particles only within the region of
  interest reduces the total mass of ejecta that must be covered (less
  than $0.03 M_\odot$) and makes it easy to use higher mass resolution
  where it is most needed, i.e., in the neutrino-processed ejecta.  \X{For a more general overview of the problems
  in extracting tracer trajectories from
  multi-D simulations, we refer the reader to
  \citet{Harris2017}.}

\subsection{Nucleosynthesis conditions in the ejecta }

Some nucleosynthesis-relevant properties for all SN models are
summarized in Table~\ref{tab:properties}.
It is important to note that we only calculate the nucleosynthesis
for the ``early ejecta'',
i.e., consider only the material
  that has been ejected in neutrino-driven outflows or undergone
  explosive burning in the shock by the end of the simulations.
  The contributions of the outer shells
  \X{(i.e., the H-He envelopes of
e8.8, z9.6, and u8.1 and material outside
the middle of the O burning shell in s11, s15, and s27)}
  and later neutrino-driven
  ejecta to the total yields are neglected. 
  \X{For the massive CCSNe (s11, s15, s27), the nucleosynthetic contribution for the outer part of the O shell therefore
  remains uncertain, and  long-term simulations
to several seconds \citep{Mueller2015,Mueller2017} will be necessary to determine it.
As the shock has only traversed part of the
O shell in these models, we also miss part
of the contribution of hydrostatic and explosive burning to heavy element production
from O/C/He shells so that our yields
can only be understood as lower limits
for some species (among them $^{26}\mathrm{Al}$ and $^{60}\mathrm{Fe}$).
We also disregard the possibility of late-time fallback
of the initial ejecta, but 
  this may not be a major uncertainty since the most advanced
parameterised 1D models of neutrino-driven
explosions suggest that this plays a minor role \citep{Ertl2016}
for the progenitor mass range considered here.}

  For models e8.8, z9.6, and u8.1, explosive burning is already
  complete at the end of the simulations:
  \X{The outer shell (composed of H and He) will not add
  significant yields of heavy elements, and 
  no more explosive
  nucleosynthesis is taking place as the shock
  propagates to the stellar envelope after the end of
  the simulations.} We still miss small amounts of
  ejected material from the neutrino-driven wind
 \X{\citep{Duncan1986}} in these progenitors,
  however. This material may undergo weak $r$- or $\nu p$-process
  nucleosynthesis, though the most recent calculations of wind
  nucleosynthesis for e8.8 suggest only a relatively unspectacular
  production of iron group elements and some $\nu p$-process nuclides
  with small production factors \citep{Pllumbi2015}, which turn out to
  be insignificant compared to the contribution from the early ejecta
  that we investigate here.  For e8.8, z9.6, and u8.1, the
    yields presented here thus cover essentially the complete
    nucleosynthesis of heavy elements in these progenitors. 

    The case is different for s11, s15, and s27, where the shock
  has progressed only to the middle of the O/Si shell. For these
  models, substantial amounts of ashes from O, Ne, C, and He burning
  thus remain to be ejected and will contribute significantly to the
  total production factors. The post-shock temperatures
  at the end of our simulations also remain sufficiently high
  for some additional explosive O burning to take place. Moreover,
  strong accretion downflows still persist in these models. This keeps
  the neutrino luminosities high and allows the ejection of
  neutrino-heated matter to continue well beyond $1 \, \mathrm{s}$
  after the onset of the explosion
  \citep{Mueller2015,Bruenn2016}. Our nucleosynthesis calculations
  therefore place only a lower bound on the production of
  iron-group and trans-iron group elements in the supernova core
  of these models. It is noteworthy that this may partly
  contribute to the unexpectedly small mass of $^{56}$Ni
  obtained for these models, which remains significantly
  smaller than expected for ``ordinary'' supernovae
  \citep[e.g., $0.07\, M_\odot$ for SN~1987A,][]{Bouchet1991}. 

  Aside from the limitation of our analysis to the
  nucleosynthesis during the first few hundred milliseconds
  after shock revival, the nucleosynthesis in our models
  remains subject to to other uncertainties: The simulations
  were conducted assuming axisymmetry (2D),
  and the  explosion energies at the end of simulations are only
  (0.3--$1.5) \times
10^{50}$~erg \citep[see Fig.~15 in][]{Janka2012}, which are either less
energetic than typical observed events \citep[several $10^{50}$~erg,
 e.g.,][]{Kasen2009,Nomoto2013} or still in a phase of steep rise, i.e., not converged to their final values. We discuss the possible repercussions
  of this in 
  {\S}~\ref{subsec:ni56} and {\S}~\ref{sec:conclusions}.

Figure~\ref{fig:yehist} shows the $Y_\mathrm{e}$ distributions in the
ejecta for all models evaluated at $T_9 = 10$, where $T_9$ is the
temperature in units of $10^9$~K. For a trajectory with the maximum
temperature of $T_\mathrm{9, max} < 10$, we show $Y_\mathrm{e}$ at $T_9
= T_\mathrm{9, max}$ instead. $\Delta M_\mathrm{ej}$ is the ejecta mass
in each $Y_\mathrm{e}$ bin with an interval of $\Delta Y_\mathrm{e} =
0.005$. The minimum and maximum values of
$Y_\mathrm{e}$ for all models are given in
Table~\ref{tab:properties} (8th and 9th columns, respectively). The $Y_\mathrm{e}$ distributions shown
here are similar to those inferred from the hydro data at late times except for
small (up to a few \%) shifts toward $Y_\mathrm{e} \approx 0.5$ as a
result of some numerical diffusion and mixing (clipping of extrema)
that suppresses the tails of the $Y_\mathrm{e}$ distribution in
the hydro at late times\footnote{In practice, we find
  this effect to be small up to the point
  where the ejecta reach a radius of around $1000 \, \mathrm{km}$
  and the radial computational grid becomes coarser.}.
We find
that low-mass models (e8.8, z9.6, and u8.1) have appreciable amounts of
neutron-rich ejecta (40--50\%; 7th column in
Table~\ref{tab:properties}). This is due to the faster growth of the shock
radii for these low-mass cases: As a result of the higher ejecta speed, less
time is available to increase the $Y_\mathrm{e}$ by neutrino processing as
material is ejected from the neutron-rich environment near the gain
radius, \X{where conditions close to equilibrium between
neutrino absorption and electron (and to some extent positron) captures
are maintained and lead to a low $Y_\mathrm{e}$}. By contrast, the bulk of the ejecta are proton-rich (96--99\%;
Table~\ref{tab:properties}) in massive models (s11, s15, and s27).
\X{Here the ejecta expand more slowly, allowing
neutrino absorption to reset the $Y_\mathrm{e}$
to an equilibrium value determined by the competition
of electron neutrino and antineutrino absorption
\citep{Qian1996,Froehlich2006}. With similar electron neutrino
and antineutrino mean energies the equilibrium value
tends to lie above $Y_\mathrm{e}=0.5$ because of the proton-neutron mass difference.
}

\begin{figure*}
\epsscale{1.0}
\plotone{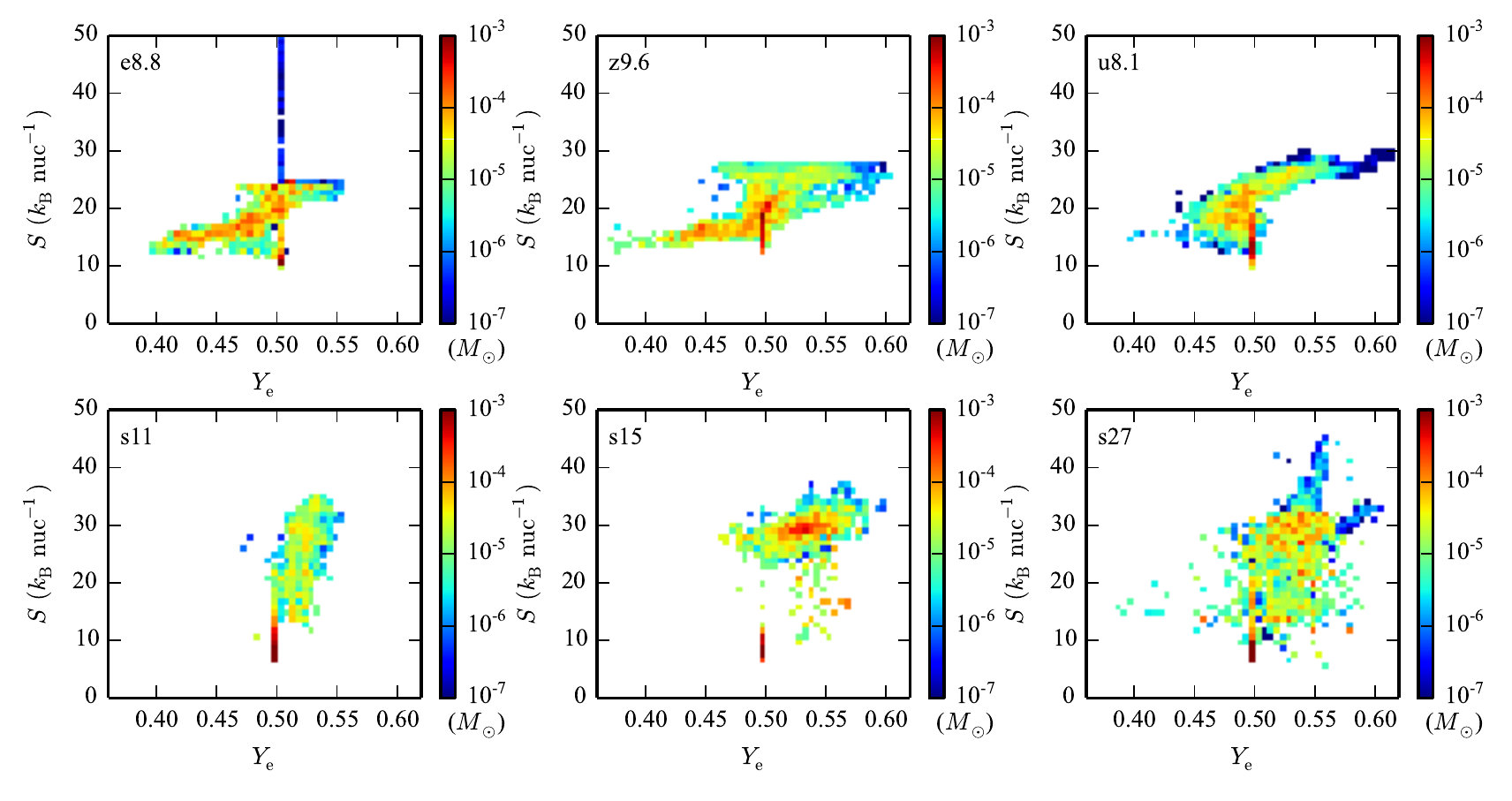}

\caption{Distributions of ejecta masses as functions of electron
fraction ($Y_\mathrm{e}$) and asymptotic entropy per nucleon ($S$) for all SN
models. Strongly neutrino-processed ejecta show a wide range of $Y_\mathrm{e}$-values
and a slightly positive correlation between $S$ and $Y_\mathrm{e}$. Material that
is shocked at relatively large radii after the onset of the explosion
appears in the form of a narrow vertical stripe with $Y_\mathrm{e} \approx 0.5$ 
in the plots and exhibits variations in $S$ depending on the
pre-shock entropy and density and the shock velocity. }
\label{fig:yes}
\end{figure*}

Figure~\ref{fig:yes} illustrates the distributions of ejecta masses as
functions of $Y_\mathrm{e}$ and asymptotic entropy per nucleon, $S$
(in units of Boltzmann's constant $k_\mathrm{B}$). All SN models show
spikes of
$Y_\mathrm{e} \approx 0.50$ components as leftovers of
the initial
composition of shocked material that never undergoes strong neutrino processing,
and 
a positive correlation of  $S$ with $Y_\mathrm{e}$ in the neutrino-processed ejecta. The latter fact is reasonable because neutrino-heating raises both
$Y_\mathrm{e}$ and $S$. We find, however, a larger scatter of the
entropies at a given $Y_\mathrm{e}$ for massive SN models (s11, s15,
and s27). This reflects the more vigorous motions arising from
convective instability and SASI for more massive cases as well as
stronger temporal and spatial variations in the neutrino irradiation.

Note that the very high entropies at
$Y_\mathrm{e} \approx 0.5$ for e8.8 (up to $S_\mathrm{max} = 383
k_\mathrm{B}$; last column in Table~\ref{tab:properties}) stem from
shock heating of outgoing material colliding with the dilute SAGB
envelope. However, a sizable post-shock  entropy, $S > 100
k_\mathrm{B}$, is reached  only in shells which never
reach 
nuclear equilibrium \citep[$T_9 < 3$,][]{Janka2008, Kuroda2008} so that these shells
do not contribute to the production of 
heavy elements \citep[such as an $r$-process suggested
by][]{Ning2007}. Except for this component of shock-heated ejecta
with low temperatures, e8.8 has a
maximum entropy of nucleosynthesis-relevant material of
$\approx 25 k_\mathrm{B}$, and more massive models
have larger values ($S_\mathrm{max}$; last column in
Table~\ref{tab:properties}).

\begin{figure*}
\epsscale{1.0}
\plotone{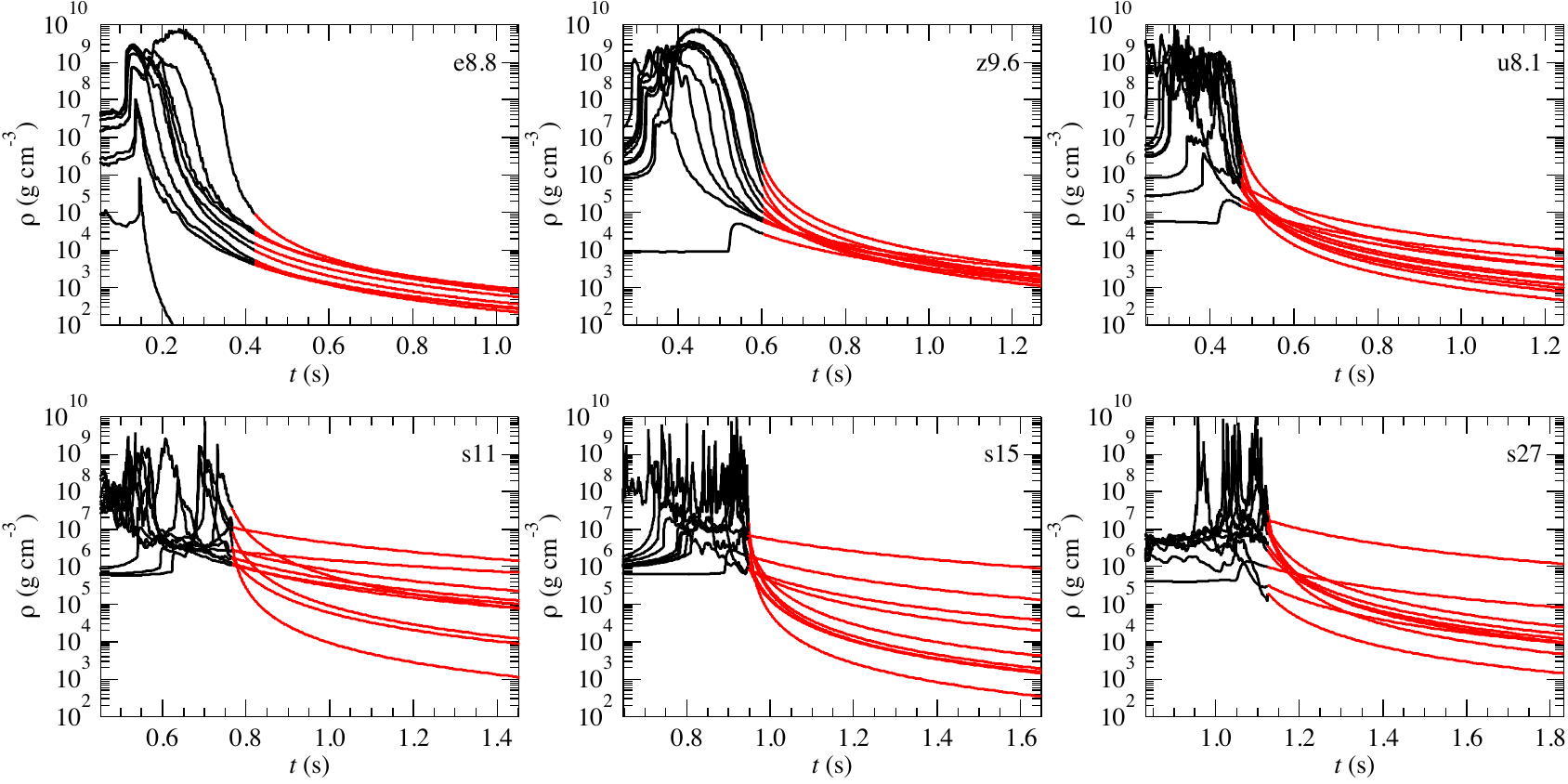}

\caption{\R{Evolutions of density as functions of times after core bounce for selected trajectories that are extrapolated (red curves) according to Eq.~(\ref{eq:density}).}}

\label{fig:trajrho}
\end{figure*}

\begin{figure*}
\epsscale{1.0}
\plotone{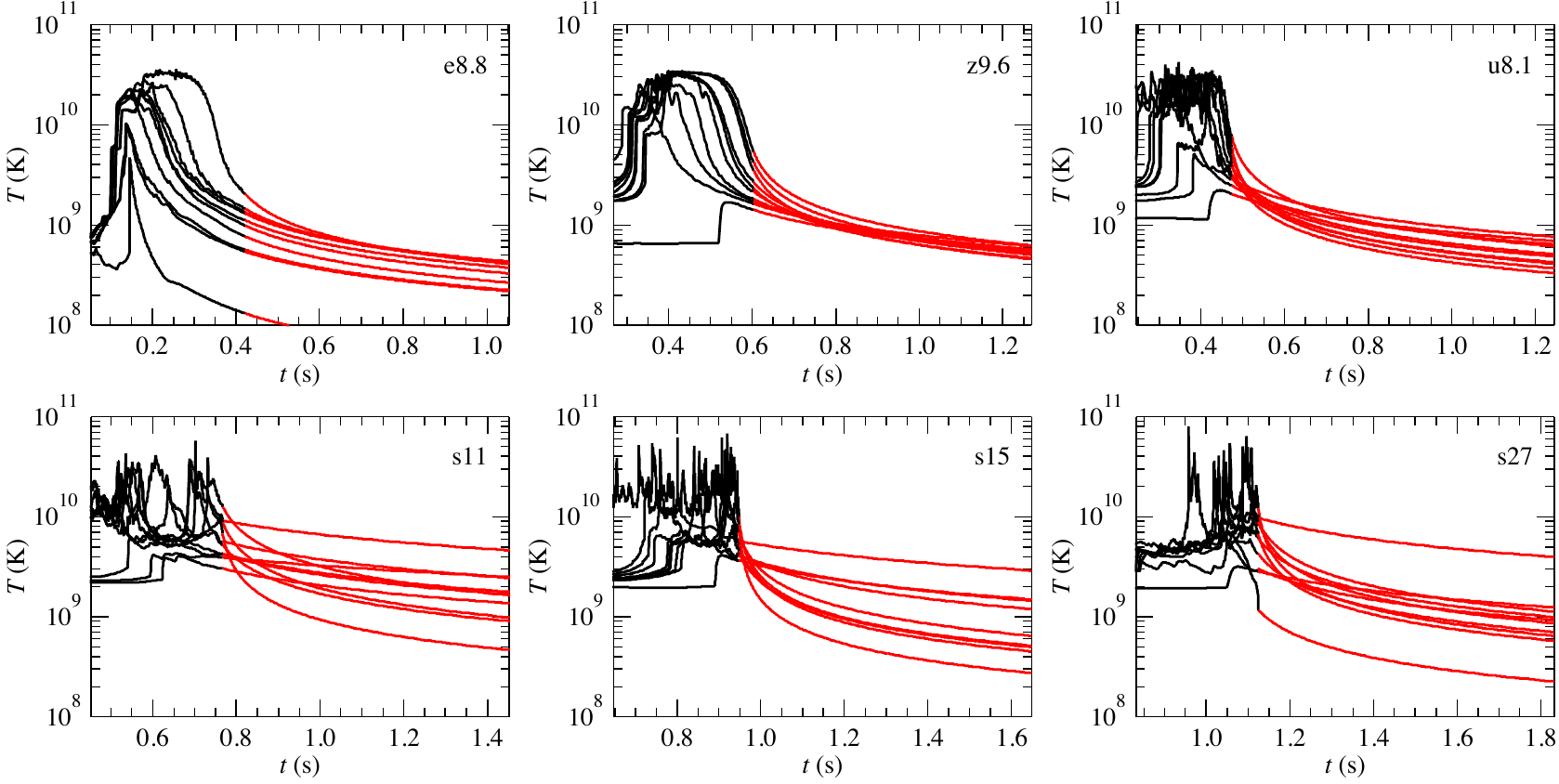}

\caption{\R{Same as Figure~\ref{fig:trajrho} but for temperature (with extrapolations according to Eq.~(\ref{eq:temperature})).}}

\label{fig:trajt}
\end{figure*}

The core-collapse simulations were stopped at $t = t_\mathrm{fin}$ for
our models, where $t_\mathrm{fin} = 0.423$~s, 0.605~s, 0.474~s,
0.767~s, 0.947~s, and 1.13~s after core bounce for e8.8, z9.6, u8.1, s11, s15, and
s27, respectively. 
At these times, the temperatures in the ejecta  were still high, in particular
for massive models, so that nucleosynthesis would be still
active. The temperature ($T$) and density ($\rho$) thus
need to be extrapolated for nucleosynthesis calculations.
It is known that the late-time evolution of the density is well
approximated by $\rho \propto t^{-2}$ \citep[e.g.,][]{Arcones2007}. We
thus assume 
\begin{eqnarray} 
\label{eq:density} \rho(t) = c_1 (t -
t_1)^{-2} \quad (t > t_\mathrm{fin}), 
\end{eqnarray} 
where the
constants $c_1$ and $t_1$ are determined to get a smooth connection of
the density at $t = t_\mathrm{fin}$. The expansion of the ejecta is almost
adiabatic at this stage (i.e., $T^3/\rho \approx
\mathrm{constant}$)\footnote{Entropy generation due to
electron-positron annihilation after the freezeout from nuclear statistical equilibrium (NSE) is not taken into account for deriving
Equation~(\ref{eq:temperature}). We expect, however, that its effect
is minor because of the strong dependencies of nucleosynthetic
abundances on $Y_\mathrm{e}$ rather than entropy or expansion
timescale.}, and thus the temperature is extrapolated such as
\begin{eqnarray} \label{eq:temperature} T(t) = c_2 (t - t_1)^{-2/3}
\quad (t > t_\mathrm{fin}) \end{eqnarray} with $t_1$ from
Equation~(\ref{eq:density}) and $c_2$ determined so that the
temperature matches the value at $t = t_\mathrm{fin}$. The radius $r$
for $t > t_\mathrm{fin}$, which is needed to calculate the rates of
neutrino interactions, is obtained from Equation~(\ref{eq:density})
with the assumption of  steady-state conditions, i.e., $r^2 \rho v_r =
\mathrm{constant}$ \citep[][]{Panov2009, Wanajo2011b}, where $v_r$ is
the radial component of velocity at $t = t_\mathrm{fin}$ and constant afterwards.
\R{Selected temporal evolutions of density and temperature are shown in Figures~\ref{fig:trajrho} and \ref{fig:trajt}, respectively, where the extrapolations are indicated by red curves. The evolutions for e8.8 and z9.6 appear to be nearly self-similar as expected from their almost spherical expansions (Fig.~\ref{fig:snimage}). For more massive models, in particular s11, s15, and s27, a variety of evolutions can be seen because of highly asymmetric, vigorous convective and SASI motions.}

\section{Nucleosynthesis}\label{sec:nucleosynthesis}

The nucleosynthesis yields in each SN trajectory are computed in a
post-processing step by solving an extensive nuclear reaction network
\citep{Wanajo2001} with the temperature and density histories
described in \S~\ref{sec:ccsnmodels}. The numbers of processed
trajectories are 2343 (e8.8), 6310 (z9.6), 4672 (u8.1), 2739 (s11),
2565 (s15), and 2312 (s27). The up-to-date network consists of 7435
isotopes between the proton- and neutron-drip lines from single
neutrons and protons up to isotopes with $Z = 110$. All the reaction
rates are taken from REACLIB
V2.0\footnote{https://groups.nscl.msu.edu/jina/reaclib/db/.}
\citep{Cyburt2010} making use of experimental data when available. As
we will see later, the nucleosynthetic abundances are mostly
determined in nuclear equilibrium in the regions relatively close to
$\beta$-stability where experimental masses are
available. Uncertainties arising from nuclear data are
thus expected to be small. Rates for electron capture
\citep{Langanke2001} as well as for neutrino interactions on free
nucleons \citep{McLaughlin1996} and $\alpha$-particles
\citep{Woosley1990} are also included. The radiation field computed
in the supernova simulations is used as input for computing
the neutrino interactions in our nucleosynthesis calculations.
We retain the \X{full dependence of the radiation field
on radius, latitude, and time}
  as computed in our ray-by-ray-plus approximation, \X{i.e., 
  we use the local neutrino energy density, number density, and
  energy moments of the distribution function at the current position
  of each tracer particle as input for the network
  calculations.}
  

Each nucleosynthesis calculation is initiated when the temperature
decreases to $T_9 = 10$ with \X{initial mass fractions 
of $Y_\mathrm{n}=1 -
Y_\mathrm{e}$ for free neutrons
and $Y_\mathrm{p}=Y_\mathrm{e}$  for free protons},
respectively\footnote{At such high temperature (and density), the
matter immediately attains nuclear statistical equilibrium (NSE) and thus any initial composition with
the total charge of $Y_\mathrm{e}$ is available.}. For the
trajectories with $T_\mathrm{9, max} < 10$, the initial compositions
adopted from our hydrodynamical simulations are utilized. 

\subsection{Types of nucleosynthesis}\label{subsec:type}

We first analyze the nucleosynthesis in detail for z9.6,
since this  model covers the widest range in $Y_\mathrm{e}$  ($Y_\mathrm{e}$=0.373--0.603; Fig.~\ref{fig:yehist}). Figure~\ref{fig:toseed} shows
the neutron-, proton-, and $\alpha$-to-seed abundance ratios
($Y_\mathrm{n}/Y_\mathrm{h}$, $Y_\mathrm{p}/Y_\mathrm{h}$, and
$Y_\alpha/Y_\mathrm{h}$) at a point when  the temperatures decrease to $T_9 = 3$
for all the trajectories of model z9.6. We consider
this temperature as it approximately corresponds to  the end of nuclear equilibrium:
This temperature is 
close to the conventional critical temperature for the termination of
quasi-nuclear equilibrium \citep[QSE; $T_9 = 4$,][]{Meyer1998b}, to the
beginning of a $\nu p$-process \citep[$T_9 = 3$,][]{Froehlich2006},
and the beginning of an $r$-process \citep[$T_9 =2.5$,][]{Woosley1994}. Here, the seed abundance $Y_\mathrm{h}$ is
defined as the total abundance of all nuclei heavier  than helium. The other SN models
show  similar dependencies of $Y_\mathrm{n}/Y_\mathrm{h}$,
$Y_\mathrm{p}/Y_\mathrm{h}$, and $Y_\alpha/Y_\mathrm{h}$ on $Y_\mathrm{e}$ (not
shown here), indicating that $Y_\mathrm{e}$ (rather than entropy and
expansion timescale) is most crucial for the nucleosynthesis in our
case. For this reason, we describe our nucleosynthetic results for
model z9.6 in terms of $Y_\mathrm{e}$. According to
Figure~\ref{fig:toseed}, we can distinguish 
 the \X{following types} of
nucleosynthesis regimes in our supernova models:
\X{Among the (strongly) neutrino-processed ejecta,
we find a} nuclear statistical equilibrium (NSE) regime for $Y_\mathrm{e}<0.43$, a quasi-nuclear equilibrium (QSE) regime
for $0.43 \le Y_\mathrm{e} < 0.5$, and nucleosynthesis by charged-particle capture
processes for $Y_\mathrm{e}\ge 0.5$.  No $r$-process is expected in our
models because of neutron-to-seed ratios far below unity over the
entire range of $Y_\mathrm{e}$ (Fig.~\ref{fig:toseed}).
\X{We note that in all of these regimes (and different
from ``classical'' explosive nucleosynthesis), the neutrino-processed
material is initially in NSE and the QSE/charged-particle capture regime is reached after an NSE phase as the temperatures
decreases during the expansion of the ejcecta.}

\X{In addition to these three regimes, there is a fourth
regime of material with $Y_\mathrm{e}\approx 0.5$ and lower entropy; this
comprises the ejecta that immediately expand after being
shocked and undergo ``classical'' explosive nucleosynthesis. 
The nucleosynthesis in this regime is not the primary
subject of this paper; for models e8.8, z9.6, and u8.1
it is unimportant in terms of yields, and for s11, s15,
and s27, we merely cover the explosive burning of a part
of the O shell, so that there is no basis for an extensive analysis of this regime. Moreover, the
physical principles of classical explosive
nucleosynthesis are well known in the literature \citep{Arnett1996}.}


\R{It is worth noting that although the regimes of
NSE/QSE are also encountered in classical explosive nucleosynthesis, the conditions for the realization
of these regimes are very different: For classical explosive burning, the nucleosynthesis
is determined by the peak temperature (which depends on the explosion energy and the
mass-radius profile of the progenitor) and, in the NSE/QSE regime, on the $Y_\mathrm{e}$ in the progenitor.
In the case of the neutrino-processed ejecta, the initial temperature
is always high enough for NSE to obtain, and the interplay of multi-dimensional fluid flow
 and weak interactions of neutrinos, electrons and positions determines
the conditions at freeze-out from NSE/QSE, i.e., $Y_\mathrm{e}$, entropy, and expansion time scale.}

\subsubsection{NSE}\label{subsubsec:nse}

\begin{figure}
\epsscale{1.0}
\plotone{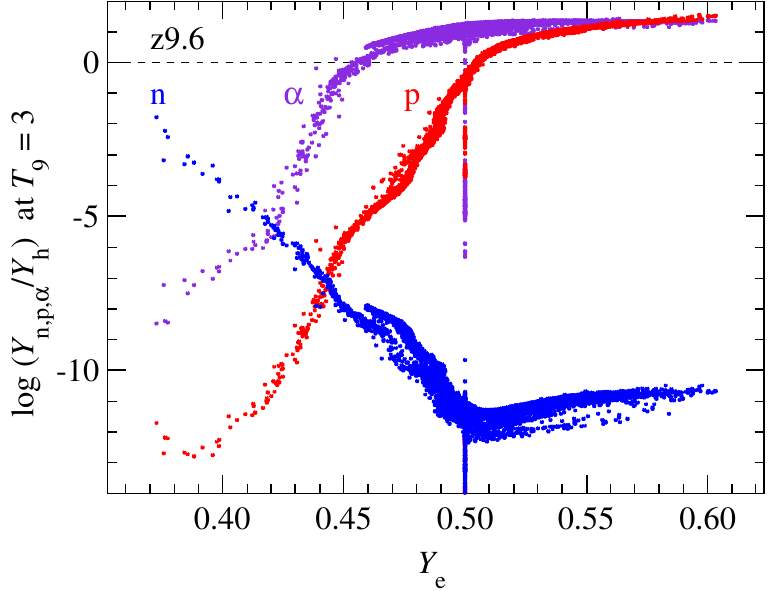}

\caption{Neutron-, proton-, and $\alpha$-to-seed abundance ratios
 (blue, red, and purple, respectively) for all trajectories of
 z9.6 at a time when  the temperatures decrease to $T_9 = 3$ as functions of
 $Y_\mathrm{e}$ (at $T_9 = 10$). }
\label{fig:toseed}
\end{figure}

For $Y_\mathrm{e} < 0.43$, all the ratios $Y_\mathrm{n}/Y_\mathrm{h}$,
$Y_\mathrm{p}/Y_\mathrm{h}$, and $Y_\alpha/Y_\mathrm{h}$ are considerably
smaller than unity ($< 0.01$) at $T_9 = 3$
(Fig.~\ref{fig:toseed}). The global distribution of final
nucleosynthetic abundances is thus mostly determined in NSE at high
temperature ($T_9 > 5$)\footnote{In this paper, we use the term
``NSE'' even for the case of $\alpha$-deficient QSE \citep{Meyer1996,
Wanajo2013a}, in which NSE serves as a reasonable guideline for
abundance determinations.}. The subsequent QSE does not substantially
change the abundance distribution because of the small amounts of free
nucleons and $\alpha$ particles. This is due to relatively small
entropies ($S \sim 14\, k_\mathrm{B}/\mathrm{nuc}$; Fig.~\ref{fig:yes})
for these neutron-rich ejecta and the neutron-richness itself. Under such
conditions the three-body process $\alpha (\alpha n, \gamma) ^9$Be
(followed by $^9$Be$(\alpha, \gamma) ^{12}$C), rather than
triple-$\alpha$, is fast enough to form the NSE cluster by assembling
free nucleons and $\alpha$ particles. Such neutron-rich conditions also
disfavor $\alpha$ emission (i.e., to avoid to be more
neutron-rich). In NSE, the resulting abundance distribution is
independent of specific reactions. What determines the abundance
distribution are the binding energies per nucleon ($B/A$; shown in
Figure~\ref{fig:bsna}, left). As the $Y_\mathrm{e}$ of nuclides
(indicated by white lines) decreases from $\sim 0.50$ to $\sim 0.40$,
the nuclides with the maximal $B/A$ shift from $^{56}$Ni ($Z = N =
28$) to $^{48}$Ca ($Z = 20$ and $N = 28$) and $^{84}$Se \citep[$Z =
34$ and $N = 50$, see][]{Hartmann1985}.

We find such NSE-like nucleosynthesis features  in
Figure~\ref{fig:freeze}, which  shows the abundance distribution in
selected trajectories of model z9.6 when the temperatures decrease to
$T_9 = 5$, 4, and 3 as well as the abundances at the end of calculations. The
top three panels correspond to the trajectories with NSE-like
conditions of $Y_\mathrm{e} < 0.43$ and relatively low entropies
($\sim 14\, k_\mathrm{B}/\mathrm{nuc}$). Major abundance peaks are
already formed at $T_9 = 5$ (red lines). The abundance patterns are
almost frozen when the temperature decreases to $T_9 = 4$ (defined as
the end of NSE) and do not change significantly during the subsequent evolution.

\subsubsection{QSE}\label{subsubsec:qse}

\begin{figure*}
\epsscale{1.0}
\plotone{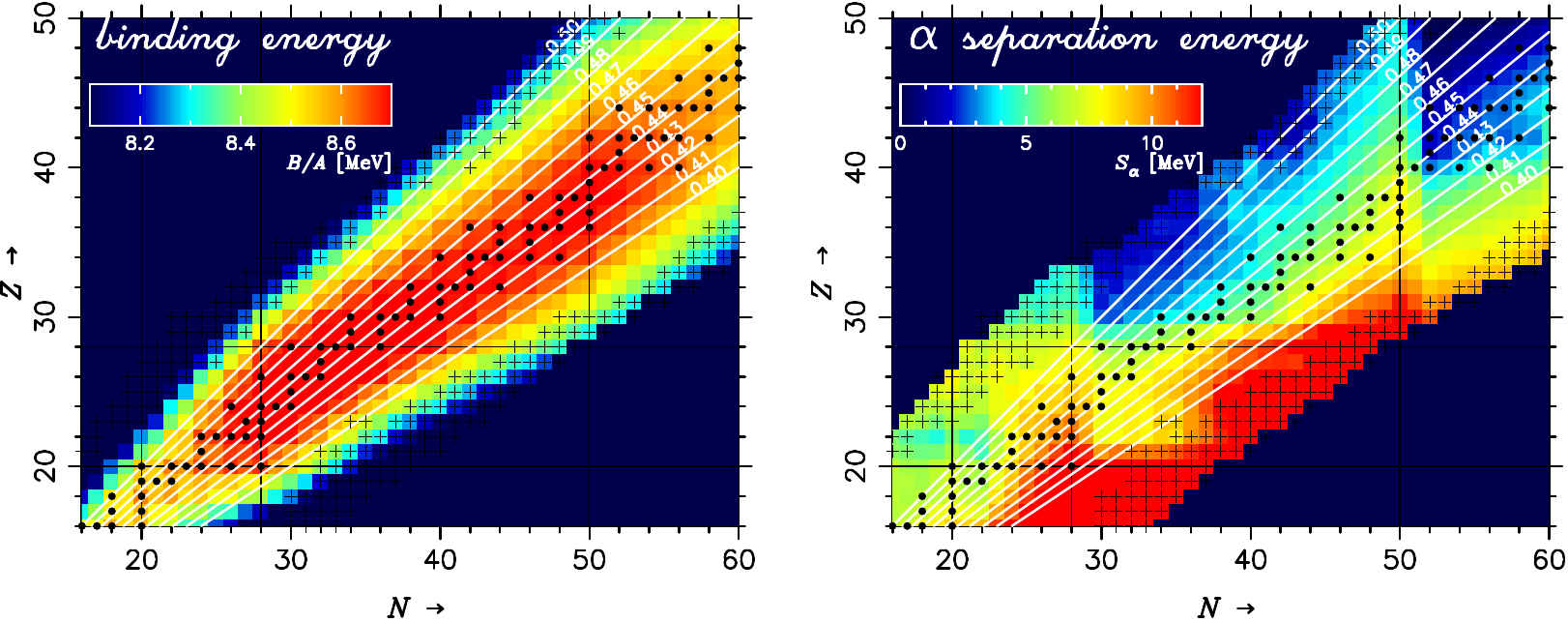}

\caption{Binding energies per nucleon (left) and $\alpha$-separation
 energies (right) in units of MeV for nuclides with experimentally
 evaluated masses \citep[crosses mark the extrapolated
 values,][]{Audi2012}. The dots denote stable isotopes. The black
 and white lines indicate neutron and proton magic numbers (20,
 28, and 50) and the $Y_\mathrm{e}$ values of nuclides, respectively.}

\label{fig:bsna}
\end{figure*}

For $0.43 \le Y_\mathrm{e} < 0.50$, the $\alpha$ concentration becomes
important ($Y_\alpha/Y_\mathrm{h} \sim 0.01$--10) but
$Y_\mathrm{p}/Y_\mathrm{h}$ is less than unity at $T_9 = 3$. In this
case, the final abundances are mainly determined in QSE, a subsequent
stage after the $\alpha$-rich freeze-out from NSE \citep{Woosley1992,
Meyer1998b}. At the end of NSE ($T_9 \sim 5$), the single NSE cluster
splits into two QSE clusters with one consisting of free nucleons and
$\alpha$ particles and the other consisting of heavy nuclei. In the
latter QSE cluster, the heavy nuclei are absorbed in the
``$\alpha$-bath'' and its distribution is determined by the $\alpha$
separation energies (shown in the right panel of
Figure~\ref{fig:bsna}), independent of specific nuclear
reactions. With a modest neutron-richness of $Y_\mathrm{e} \sim
0.43$--0.49 (values are indicated by white lines in
Fig.~\ref{fig:bsna}), the nuclides near $N= 28$ and 50 such as
$^{64}$Zn, $^{88}$Sr, $^{89}$Y, $^{90}$Zr, and $^{92}$Mo are
preferentially formed in QSE owing to their greater $\alpha$
separation energies \citep[see also][]{Hoffman1996, Wanajo2006}.

We find in the left-middle and middle panels of
Figure~\ref{fig:freeze} ($Y_\mathrm{e} = 0.450$ and 0.475,
respectively) that the abundance distributions substantially change
during QSE ($T_9 \sim 5$ to 4). The abundance patterns are determined
roughly when  the temperature decreases to $T_9 = 3$, and further evolution
are unimportant (although an enhancement of nuclei with $A = 10$--50 can be seen as a result of $\alpha$-particle capture).

\subsubsection{Charged-particle capture process}\label{subsubsec:ccap}

\begin{figure*}
\epsscale{1.0}
\plotone{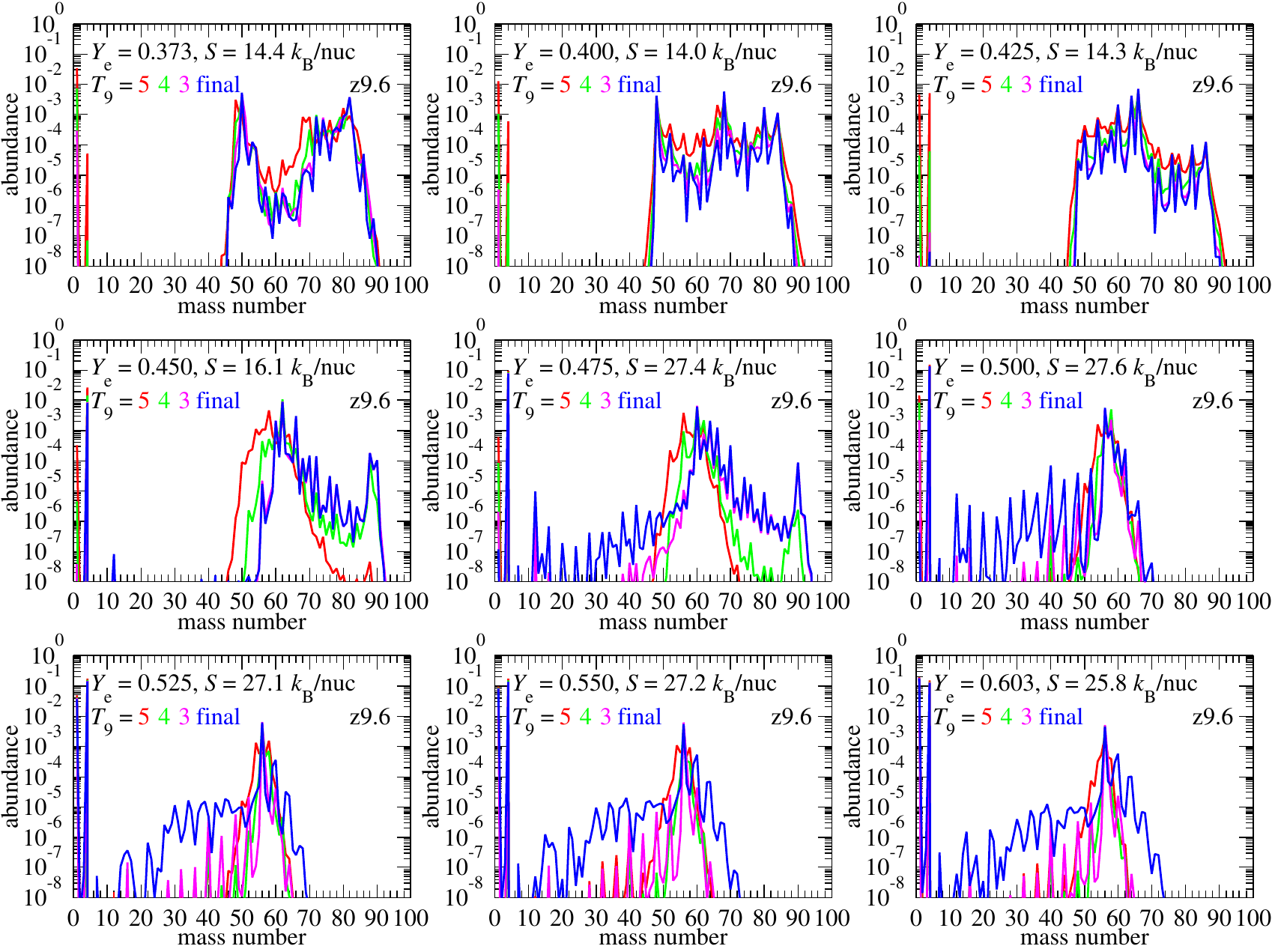}

\caption{Nucleosynthetic abundances for model z9.6 when the
 temperatures decrease to $T_9 = 5$, 4, and 3 as well as those at the
 end of calculations (``final''). Selected trajectories are those with
 initial $Y_\mathrm{e} = 0.373$ (left-top), 0.400 (middle-top), 0.425
 (right-top), 0.450 (left-middle), 0.475 (middle), 0.500
 (right-middle), 0.525 (left-bottom), 0.550 (middle-bottom), and 0.603
 (right-bottom). The asymptotic entropy $S$ is also shown in the
 legend of each panel.}

\label{fig:freeze}
\end{figure*}

For $Y_\mathrm{e} \ge 0.50$, $^{56}$Ni dominates in the heavy QSE
cluster, and the $\alpha$-rich freeze-out from QSE ($T_9 \sim 4$) leads
to an $\alpha$-process \citep{Woosley1992}. This greatly enhances the
abundances of $\alpha$-elements with $A \sim 12$--40 (with multiples
of 4) as can be seen in the right-middle panel of
Figure~\ref{fig:freeze} (for $Y_\mathrm{e} = 0.500$). For
$Y_\mathrm{e} > 0.51$, both $Y_\alpha/Y_\mathrm{h}$ and
$Y_\mathrm{p}/Y_\mathrm{h}$ are greater than unity at $T_9 = 3$
(Fig.~\ref{fig:toseed}). The freeze-out is thus followed by
$\alpha$-capture and proton-capture processes. We find in the bottom
three panels of Figure~\ref{fig:freeze} ($Y_\mathrm{e} = 0.525$,
0.550, and 0.603) that nuclei in a wide range of $A\sim 10$--70,
including odd-$Z$ elements, are substantially enhanced by these
charged-particle capture processes after the temperature drops below
$T_9 = 3$.

\begin{figure}
\epsscale{1.0}
\plotone{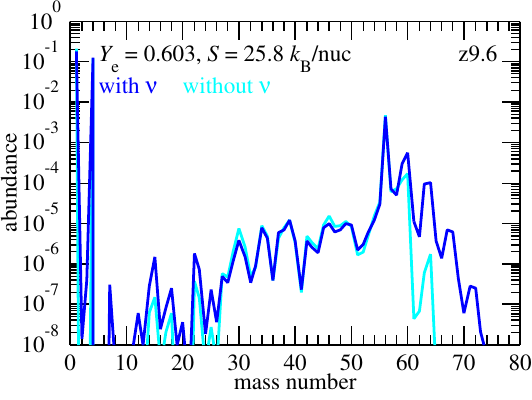}

\caption{Same as the final abundance pattern in the right-bottom panel
of Figure~\ref{fig:freeze}, but 
neglecting neutrino interactions below $T_9=10$ 
(cyan).}

\label{fig:nuonoff}
\end{figure}

Figure~\ref{fig:nuonoff} compares the final abundances with (blue) and
without (cyan) neutrino reactions for the trajectory that has the
highest $Y_\mathrm{e} = 0.603$ in model z9.6 (same as that in the
right-bottom panel of Fig.~\ref{fig:freeze}). This indicates that the
enhancement of nuclei with $A = 60$--70 is due to a $\nu p$-process, in
which the faster $(n, p)$ and $(n, \gamma)$ reactions with the free
neutrons supplied by $\bar{\nu}_e$ capture on free protons replace the
slower $\beta^+$-decays \citep{Froehlich2006, Pruet2006,
Wanajo2006}. Note that the $\nu p$-process in our result is very weak
and only the nuclei up to $A \sim 70$ are produced despite its
substantial proton-richness (up to $Y_\mathrm{e} = 0.603$). By contrast,
\citet{Wanajo2011b} have shown that the $\nu p$-process in
neutrino-driven wind with similar proton-richness can produce nuclei up
to $A \sim 120$ \citep[see also][]{Pruet2006, Arcones2012}. This
discrepancy is due to the lower entropies and longer expansion
timescales in the early dynamical ejecta, which reduces
$Y_\mathrm{p}/Y_\mathrm{h}$ at the beginning of a $\nu p$-process ($T_9
\sim 3$), than those in the late-time neutrino-driven wind.

\subsection{Dependencies of isotope productions on $Y_\mathrm{e}$}\label{subsec:isotope}

In \S~\ref{subsec:type} we found that the iron-group (from Ca to Cu)
and light trans-iron (from Zn to Mo) species can be produced in the
innermost ejecta of z9.6, the model with the widest
range in $Y_\mathrm{e}$ in the ejecta . No
heavier elements are produced in any of our six  models. In
Figures~\ref{fig:yeabun1} and \ref{fig:yeabun2}, the final mass
fractions of stable isotopes from K to Mo are presented as functions
of $Y_\mathrm{e}$ for all the trajectories of model z9.6. The mass
fractions of selected radioactive isotopes (before decay) are also
shown in the right-bottom panel of Figure~\ref{fig:yeabun2}. We find
from these figures that few isotopes exhibit maximum abundances near
$Y_\mathrm{e} = 0.5$, in particular the light trans-iron species (from
Zn to Mo). Overall, Ni and light trans-iron species are predominantly
formed in neutron-rich ejecta, although the weak $\nu p$-process in
proton-rich ejecta plays a sub-dominant role for those up to
Ge. Several isotopes such as $^{48}$Ca, $^{50}$Ti, $^{54}$Cr, and
radioactive nuclide $^{60}$Fe are made only in very neutron-rich
ejecta with $Y_\mathrm{e} \sim 0.40$--0.43 \citep{Wanajo2013a,
Wanajo2013b}. This sensitivity  clearly demonstrates the importance of
nucleosynthesis studies based on multi-dimensional SN simulations
with detailed multi-group neutrino transport, which are
indispensable for accurately determining the $Y_\mathrm{e}$-distribution in the ejecta. 

\section{\X{Possible Contribution to the Solar Abundance Distribution}}
\label{sec:gce}

For each SN model, we calculate  mass-integrated abundances from
the nucleosynthetic outcomes of all trajectories.
In the following subsections,
we discuss the 
possible contribution of products from the innermost ejecta of SNe 
\X{with different explosion dynamics during the first
second after collapse} to
the Galaxy by comparing our nucleosynthesis yields to
the solar values. In particular, we focus on the production of several key
species such as $^{48}$Ca, Zn isotopes, light trans-iron elements,
$p$-nuclide $^{92}$Mo, and the radioactive isotopes $^{56}$Ni and
$^{60}$Fe. \X{It is important to stress that our goal here
is not to predict the full core-collapse supernova nucleosynthesis
across the whole range of progenitors, which would require
both longer simulations and a larger grid of models at
different metallicities. Even with our small set
of models, we can, however, already
establish whether the innermost ejecta 
can provide
an important contribution to the solar inventory for plausible
rates of the different explosion behaviours exemplified
by our six models. It is particularly noteworthy that focusing
on the neutrino-processed ejecta allows us to neglect the influence
of the progenitor metallicity on the yields, since any memory
of the progenitor composition is erased in this ejecta component;
i.e., the production of heavy elements in the neutrino-heated ejecta is
always a primary process.
}

\subsection{Comparison with the elemental solar abundance}\label{subsec:esolar}

For all SN models, the mass-integrated yields are compared to the
solar abundance \citep{Lodders2003}. Figure~\ref{fig:pfel} shows the
elemental mass fractions in the total ejecta with respect to their
solar values \R{(not to the initial compositions of progenitors)}, i.e., ``production factors", for these models. The total
ejecta mass from each SN is taken to be the sum of the ejected mass
from the core and the outer envelope, that is, $M_\mathrm{prog} -
M_\mathrm{PNS}$ in Tables~\ref{tab:models} and \ref{tab:properties}. Here, the outer
envelope is assumed to be metal-free (i.e., H and He only), which is
reasonable for the models near the low-mass end of the SN range, e8.8,
z9.6, and u8.1:
\X{The part of the H and He shells that were not included in our nucleosynthesis
calculations would not contribute substantially to the yields
for the range of nuclei considered here ($Z\geq 19$) regardless
of progenitor metallicity.}
For the more massive models 
s11, s15, and s27, one should bear in mind
that there is an important additional contribution to heavy elements
(those heavier than helium) from the outer envelope \citep[see][and
references therein]{Woosley2002,Woosley2007}.

In each panel of Figure~\ref{fig:pfel}, we show a ``normalization band''
in yellow, which covers the range within $1 \, \mathrm{dex}$ of the maximum production factor and
one-tenth of that. The elements that reside in
this band can be (at least in part) originate from SNe represented by
each model, provided that their production factors are greater than
$\sim 10$ \cite[e.g.,][]{Woosley2007}. As an order-of-magnitude estimate, the
production factors of $\sim 40$ for light trans-iron elements from Zn
to Zr in our representative model z9.6 suggest that such low-mass
CCSNe can be the major sources of these elements if these events
account for a few 10\% of all CCSNe and would still contribute
sizable amounts if they make up $\mathord{\sim} 10 \%$
of all CCSNe.
This suggests that SNe at 
the low-mass end of the progenitor spectrum  as represented by models e8.8, z9.6, and u8.1 could be
an important source of light trans-iron elements
(\S~\ref{subsec:transiron}).

Note that there is a remarkable agreement  of the nucleosynthesis of
z9.6 with that of e8.8, which is a consequence of the similarity of their
pre-SN core structures with a steep density gradient outside
the core and a very dilute
outer envelope. In both cases, the  resulting fast explosion lead to the
ejection of appreciable amounts of neutron-rich material. Model u8.1
with a core structure very similar to (but with a slightly shallower
density gradient than) that of z9.6 results in, however, a different
nucleosynthetic trend with deficiencies of several elements between Zn
and Zr. This indicates that only a slight difference of pre-SN
core-density structures can lead to substantially different
nucleosynthesis outcomes.

For more massive models, the production factors are smaller than $\sim
10$ for all elements, despite their greater ejecta masses compared
to low-mass models ($M_\mathrm{ej}$ in
Table~\ref{tab:properties}). The maximum production factor of $\sim 1$
for s11 gives no element to be responsible for the Galactic chemical
evolution. For s15 and s27, only Zr with the production factor $\sim
10$ is the elements that can be originated from these types of
CCSNe. These SNe can be, however, possible contributors of some
isotopes as discussed in \S~\ref{subsec:isolar}. Moreover,  the non-neutrino processed
``outer ejecta'' from these stars, which are not fully included in our study, contribute significantly
to the chemical enrichment of the Galaxy as discussed before.  

\subsection{Comparison with the isotopic solar abundance}\label{subsec:isolar}

\begin{figure*}
\epsscale{1.0}
\plotone{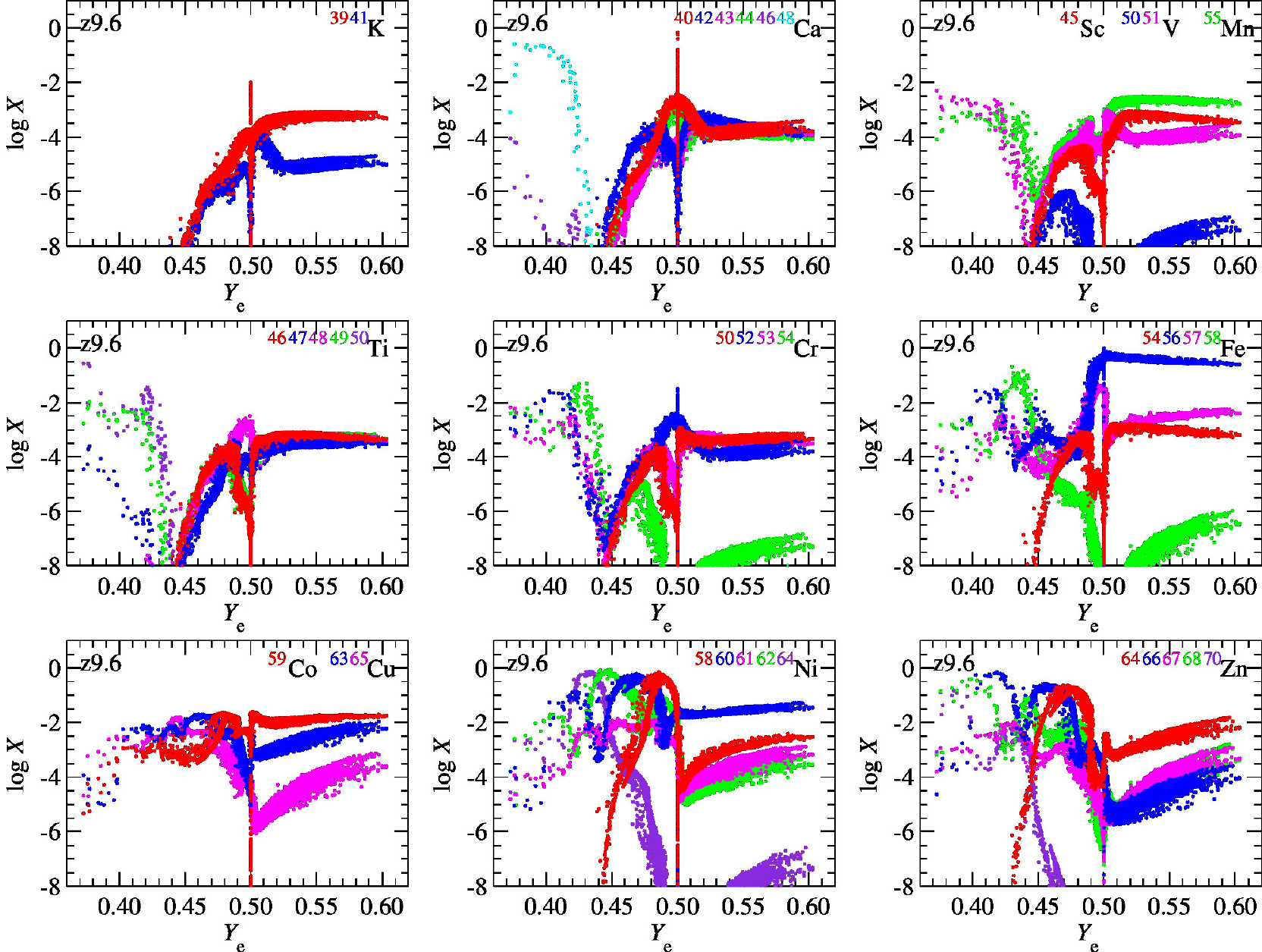}

\caption{Final mass fractions (after $\beta$-decay) of the stable isotopes of K
 (left-top), Ca (middle-top), Sc, V, and Mn (right-top), Ti
 (left-middle), Cr (middle), Fe (right-middle), Co and Cu
 (left-bottom), Ni (middle-bottom), and Zn (right-bottom) as functions
 of $Y_\mathrm{e}$ for all the trajectories of model z9.6.}

\label{fig:yeabun1}
\end{figure*}

\begin{figure*}
\epsscale{1.0}
\plotone{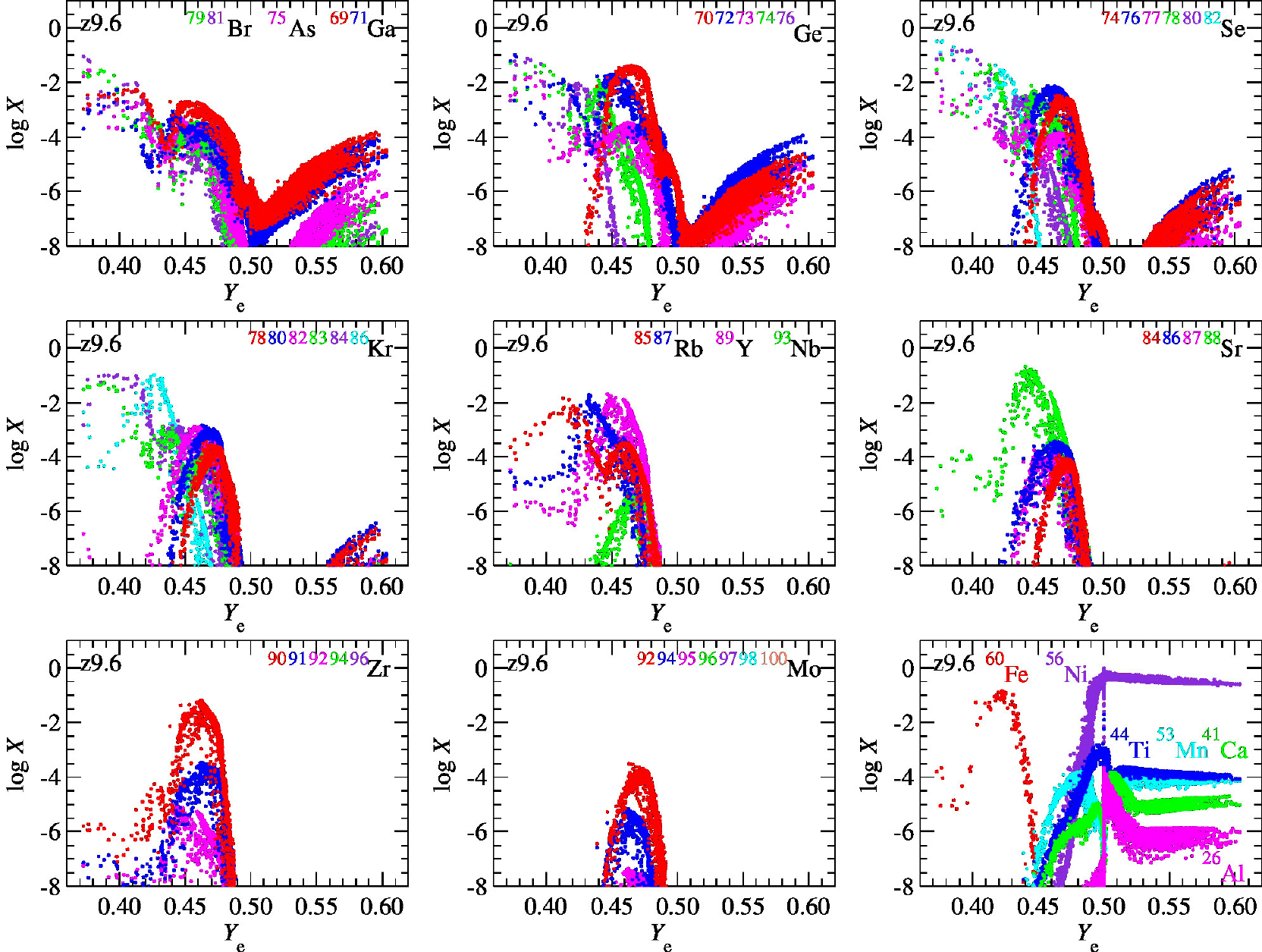}

\caption{Same as Figure~\ref{fig:yeabun1}, but for the stable isotopes of Br, As, and Ga
(left-top), Ge (middle-top), Se (right-top), Kr (left-middle), Rb, Y,
and Nb (middle), Sr (right-middle), Zr (left-bottom), Mo
(middle-bottom), and the radioactive isotopes $^{26}$Al, $^{41}$Ca,
$^{44}$Ti, $^{53}$Mn, $^{60}$Fe, and $^{56}$Ni (right-bottom).}

\label{fig:yeabun2}
\end{figure*}

Figure~\ref{fig:pfiso} compares the isotopic abundances with the solar
values for all SN models. The maximum isotopic production factor for
each model is generally greater than that of elements
(\S~\ref{subsec:esolar}), and therefore places tighter constraints on 
the contribution of relevant SNe to the Galaxy. In
Table~3, we list the 5 largest production factors for each model. Note that most of the isotopes listed here are made in
nuclear equilibrium, and thus uncertainties in individual nuclear
reaction rates are irrelevant. Models e8.8 and z9.6 exhibit
appreciable production factors of 354 ($^{86}$Kr) and 168 ($^{82}$Se),
respectively. This implies that such low-mass SNe account for $\sim
10\%$ of all CCSN events (according to the reason described in
\S~\ref{subsec:esolar}), 
supposing that these trans-iron elements originate solely
from this class of events. 
A similar contribution of 
u8.1-like events 
can be expected with its largest production
factor of 149 ($^{74}$Se). The maximum  production factor of 1.48 for s11 indicates
that no species are dominantly produced in the innermost
ejecta of  such SNe. For s15 and s27,
the largest production factors are 24.8 and 43.1 ($^{74}$Se),
respectively, which are sizably greater than those of elements ($\sim
10$, Fig.~\ref{fig:pfiso}). This indicates that such intermediate-mass
and massive CCSNe can be important  sources of several species listed in
Table~3.

To be more quantitative, we consider $^{82}$Se with the largest
production factor for model z9.6 as representative. By assuming
$f_\mathrm{z9.6}$ to be the fraction of z9.6-like events to all CCSNe,
we have \citep{Wanajo2011a} \begin{equation}
\frac{f_\mathrm{z9.6}}{1-f_\mathrm{z9.6}} =
\frac{X_\odot(^{82}\mathrm{Se})/X_\odot(^{16}\mathrm{O})}
{M_\mathrm{z9.6}(^{82}\mathrm{Se})/\langle M(^{16}\mathrm{O})\rangle}
= 0.164, \label{eq:frac}\end{equation} where
$X_\odot(^{82}\mathrm{Se}) = 1.38 \times 10^{-8}$ and
$X_\odot(^{16}\mathrm{O}) = 6.60 \times 10^{-3}$ are the mass
fractions in the solar system \citep{Lodders2003},
$M_\mathrm{z9.6}(^{82}\mathrm{Se}) = 1.91 \times 10^{-5}\, M_\odot$ is
the ejecta mass of $^{82}$Se for z9.6, and $\langle
M(^{16}\mathrm{O})\rangle = 1.5\, M_\odot$ is the production of
$^{16}$O by massive CCSNe averaged over the stellar initial mass
function between $13\, M_\odot$ and $40\, M_\odot$
\citep[see][]{Wanajo2009}.
Equation~(\ref{eq:frac}) gives
$f_\mathrm{z9.6} = 0.14$, which corresponds to a mass window $\Delta
M_\mathrm{prog} \sim 1\, M_\odot$ near the low-mass end of the SN progenitor spectrum,
$M_\mathrm{prog} \sim 9\, M_\odot$. It is currently uncertain whether
this mass window for z9.6-like progenitors is reasonable or not, as
only a slight difference of core-density structure leads to a very
different nucleosynthetic result of u8.1. If we apply
Equation~(\ref{eq:frac}) to model e8.8 by replacing $^{82}$Se with
$^{86}$Kr (Table~3), we get $f_\mathrm{e8.8} =
0.085$. This corresponds to a mass window of $\Delta M_\mathrm{prog}
\sim 0.5\, M_\odot$, which is in reasonable agreement with the
prediction from synthetic SAGB models at solar
metallicity \citep{Poelarends2008}. 
\X{If other production channels were to contribute significantly
to these elements, the allowed rate of ECSNe or z9.6-like explosion
would, of course, be lower. }

It is thus well conceivable that a
combination (or either) of ECSNe and low-mass CCSNe accounts for the
production of light trans-iron species in the Galaxy.
For the more massive
models s11, s15, and s27, a lack of self-consistent nucleosynthesis
yields in the outer envelopes precludes a quantitative estimate of
their contributions.

.

\subsection{Neutron-rich vs. proton-rich ejecta}\label{subsec:nprich}

\begin{deluxetable*}{ccccccccccc}
\tabletypesize{\scriptsize}
\tablecaption{Top 5 Production Factors}
\tablewidth{0pt}
\tablehead{
\colhead{Model} &
\colhead{1} &
\colhead{} &
\colhead{2} &
\colhead{} &
\colhead{3} &
\colhead{} &
\colhead{4} &
\colhead{} &
\colhead{5} &
\colhead{} 
}
\startdata
e8.8 & $^{86}$Kr & 354 & $^{87}$Rb & 260 & $^{82}$Se & 252 & $^{74}$Se & 222 & $^{88}$Sr & 210 \\
z9.6 & $^{82}$Se & 168 & $^{88}$Sr & 166 & $^{90}$Zr & 123 & $^{74}$Se & 110 & $^{87}$Rb & 102 \\
u8.1 & $^{74}$Se & 149 & $^{90}$Zr & 82.6 & $^{70}$Ge & 47.9 & $^{78}$Kr & 38.9 & $^{64}$Zn & 33.8 \\
s11 & $^{64}$Zn & 1.48 & $^{45}$Sc & 1.01 & $^{60}$Ni & 0.857 & $^{32}$S & 0.743 & $^{40}$Ca & 0.632 \\
s15 & $^{74}$Se & 24.8 & $^{90}$Zr & 15.4 & $^{78}$Kr & 6.97 & $^{64}$Zn & 6.10 & $^{70}$Ge & 5.36 \\
s27 & $^{74}$Se & 43.1 & $^{78}$Kr & 35.4 & $^{92}$Mo & 30.7 & $^{90}$Zr & 17.2 & $^{84}$Sr & 12.5\\
\enddata

\label{tab:pfac}

\end{deluxetable*}

Our results described in \S~\ref{subsec:esolar} and \ref{subsec:isolar}
shows that SNe near the low-mass end produce appreciable amounts of
trans-iron species despite their small ejecta masses (compared to
those of massive models; $M_\mathrm{ej}$ in
Table~\ref{tab:properties}). The reason can be found in
Figure~\ref{fig:npiso}, which shows the fraction of the
  ejecta that originates in neutron-rich
matter ($Y_\mathrm{e} < 0.4975$, $M_\mathrm{ej, n}$ in
Table~\ref{tab:properties}) for a  given
species. For low-mass models e8.8, z9.6, and u8.1, the light
trans-iron isotopes of $A = 64$--90 are almost exclusively produced in
neutron-rich ejecta, whereas proton-rich matter plays a minor role (see
also Figs.~\ref{fig:yeabun1} and \ref{fig:yeabun2}). The subdominant
roles of  SNe from massive progenitors (represented by s11, s15, and s27) to production
of these species can be understood as a result of the small
mass of neutron-rich ejecta
for these stars  ($M_\mathrm{ej, n}$ in Table~\ref{tab:properties}
and Fig.~\ref{fig:yehist}).

\begin{figure*}
\epsscale{1.0}
\plotone{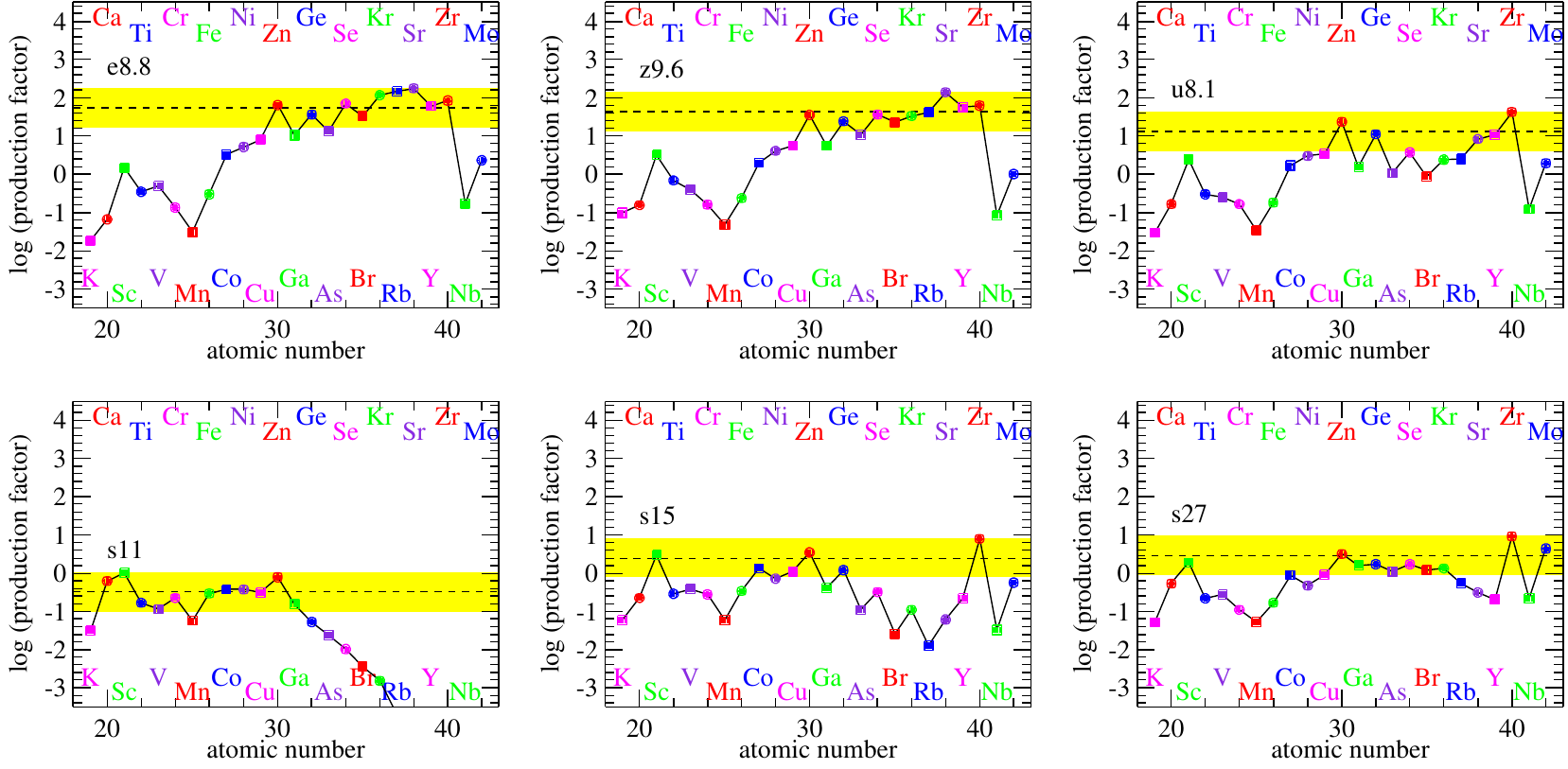}

\caption{Elemental mass fractions in the total ejecta relative to
their solar values \citep{Lodders2003}, or production factors, for all
SN models. In each panel, the normalization band, which is defined as
the range between the maximum value and one-tenth of that, is
indicated in yellow with the median value (dashed line).}

\label{fig:pfel}
\end{figure*}

Among massive models, s27 has a relatively larger amount of
neutron-rich ejecta compared to the two others, with $Y_\mathrm{e}$ down to $\sim 0.4$. This
is a consequence of the fact that model s27 exhibits an earlier explosion
with a rapidly increasing shock radius because of prominent SASI
activity that is absent in other models. These neutron-rich ejecta
lead to a relatively flat trend of production factors in this model,
similar to what we found for  e8.8 and z9.6 (Figs.~\ref{fig:pfel} and
\ref{fig:pfiso}). The bottom-panel of Figure~\ref{fig:pfiso}
indicates, however, that about a half of species between $A = 64$ and 90
originate from neutron-rich and proton-rich ejecta. In fact, 43\%
of $^{64}$Zn in s27 comes from the proton-rich ejecta. As described in
\ref{subsubsec:ccap}, a weak $\nu p$-process is responsible for the
production of these isotopes in the proton-rich ejecta. Greater
entropies of the ejecta (Fig.~\ref{fig:yes}), slower expansion
compared to those for low-mass models, and higher neutrino
luminosities and mean energies  
also enhance the efficiency of
the $\nu p$-process.

\subsection{$^{48}$Ca}\label{subsec:ca48}

Models e8.8 and z9.6 exhibit appreciable production factors of
$^{48}$Ca (18.9 and 18.6, respectively), which is made in neutron-rich
NSE \citep[or $\alpha$-deficient QSE,][]{Meyer1996, Wanajo2013a} as
can be seen in Figures~\ref{fig:freeze} and \ref{fig:yeabun1}
(middle-top panels, $Y_\mathrm{e} \sim 0.4$). The amounts are still
not large enough to regard these low-mass SNe as the main contributors
of $^{48}$Ca, although a reduction of the entropies by about
30\%  would lift
the values to a satisfactory level \citep{Wanajo2013a}. A rare class
of high-density SNe~Ia, in which similar physical conditions
are expected, has also been suggested as a possible source of
$^{48}$Ca \citep{Woosley1997}.

\subsection{Zn isotopes}\label{subsec:zn}

\begin{deluxetable*}{cccccccc}
\tabletypesize{\scriptsize}
\tablecaption{Masses of Radioactive Isotopes in the Innermost Ejecta ($M_\odot$)}
\tablewidth{0pt}
\tablehead{
\colhead{Model} &
\colhead{$^{26}$Al} &
\colhead{$^{41}$Ca} &
\colhead{$^{44}$Ti} &
\colhead{$^{53}$Mn} &
\colhead{$^{60}$Fe} &
\colhead{$^{56}$Ni} &
\colhead{$^{57}$Ni} 
}
\startdata
e8.8 & 4.39E$-$08 & 1.96E$-$07 & 2.06E$-$06 & 1.11E$-$06 & 3.61E$-$05 & 2.93E$-$03 & 1.01E$-$04\\
z9.6 & 1.29E$-$07 & 1.16E$-$07 & 2.41E$-$06 & 1.61E$-$06 & 3.14E$-$05 & 2.51E$-$03 & 9.15E$-$05\\
u8.1 & 4.77E$-$08 & 5.77E$-$08 & 1.97E$-$06 & 1.45E$-$06 & 6.65E$-$07 & 1.60E$-$03 & 7.33E$-$05\\
s11\tablenotemark{a} &  1.22E$-$09 & 1.50E$-$07 & 1.19E$-$06 & 2.53E$-$06 & 1.96E$-$20 & 3.86E$-$03 & 8.75E$-$05\\
s15\tablenotemark{a}  & 1.07E$-$08 & 2.25E$-$07 & 1.71E$-$06 & 4.46E$-$06 & 2.44E$-$18 & 6.05E$-$03 & 1.16E$-$04\\
s27\tablenotemark{a}  & 2.53E$-$08 & 4.56E$-$07 & 3.31E$-$06 & 5.89E$-$06 & 1.14E$-$06 & 5.57E$-$03 & 1.49E$-$04\\
\enddata
\tablenotetext{a}{Values for these models should be taken as lower limits (see text).}
\label{tab:radio}
\end{deluxetable*}

One of the outstanding features of our nucleosynthesis result is
production of all the stable isotopes of Zn ($A = 64$, 66, 67, 68, and
70), an element whose origin remains a mystery. Among our explored
models, the low-mass models e8.8 and z9.6 exhibit nearly flat
production factors over $A = 64$--70. This fact implies that the
element Zn, or all their stable isotopes, could  originate from ECSNe or
low-mass CCSNe as represented by e8.8 and z9.6, respectively. As can be
seen in the right-bottom panel of Figure~\ref{fig:yeabun1}, Zn
isotopes are predominantly made in neutron-rich ejecta with
$Y_\mathrm{e} \sim 0.4$--0.5; and models e8.8 and z9.6 produce a
sufficient amount of ejecta in this range to contribute most
of the solar inventory of Zn. 
There is still a
discrepancy of a factor of ten between the largest ($^{66}$Zn) and
smallest ($^{67}$Zn) production factors, which might be cured if the
$Y_\mathrm{e}$ distributions for these models were slightly
modified. \X{Alternatively, some of the neutron-rich Zn isotopes may originate from hydrostatic neutron-capture processes in massive stars \citep{Woosley1995, Nomoto2013}.} CCSNe represented by u8.1 cannot be the single source of all
Zn isotopes because of the descending trend of production factors
(Fig.~\ref{fig:pfzn}). Model s27, which shows a flat trend of production
factors, cannot be representative of Zn contributors either because
of the small production factors (1.2--5.2).

To date, only hypernova models \citep{Umeda2002, Tominaga2007,
Nomoto2013} have been proposed as sources of \X{$^{64}$Zn} \citep[except for an
  ECSN model in][]{Wanajo2011a}\X{, the dominant isotope of Zn in the solar system}. In their hypernova models 
high entropies of
ejecta lead to strong $\alpha$-rich freeze-out from nuclear
equilibrium, resulting in an appreciable production of $^{64}$Zn. This
mechanism, however,
does not co-produce 
the other
isotopes \X{and requires additional contributors such as hydrostatic neutron captures in massive stars}. 
As described in \S~\ref{subsubsec:ccap}, the $\nu
p$-process also produces $^{64}$Zn only, and therefore  cannot be the main
mechanism responsible for making  Zn.

\subsection{Light trans-iron elements}\label{subsec:transiron}

\begin{figure*}
\epsscale{1.0}
\plotone{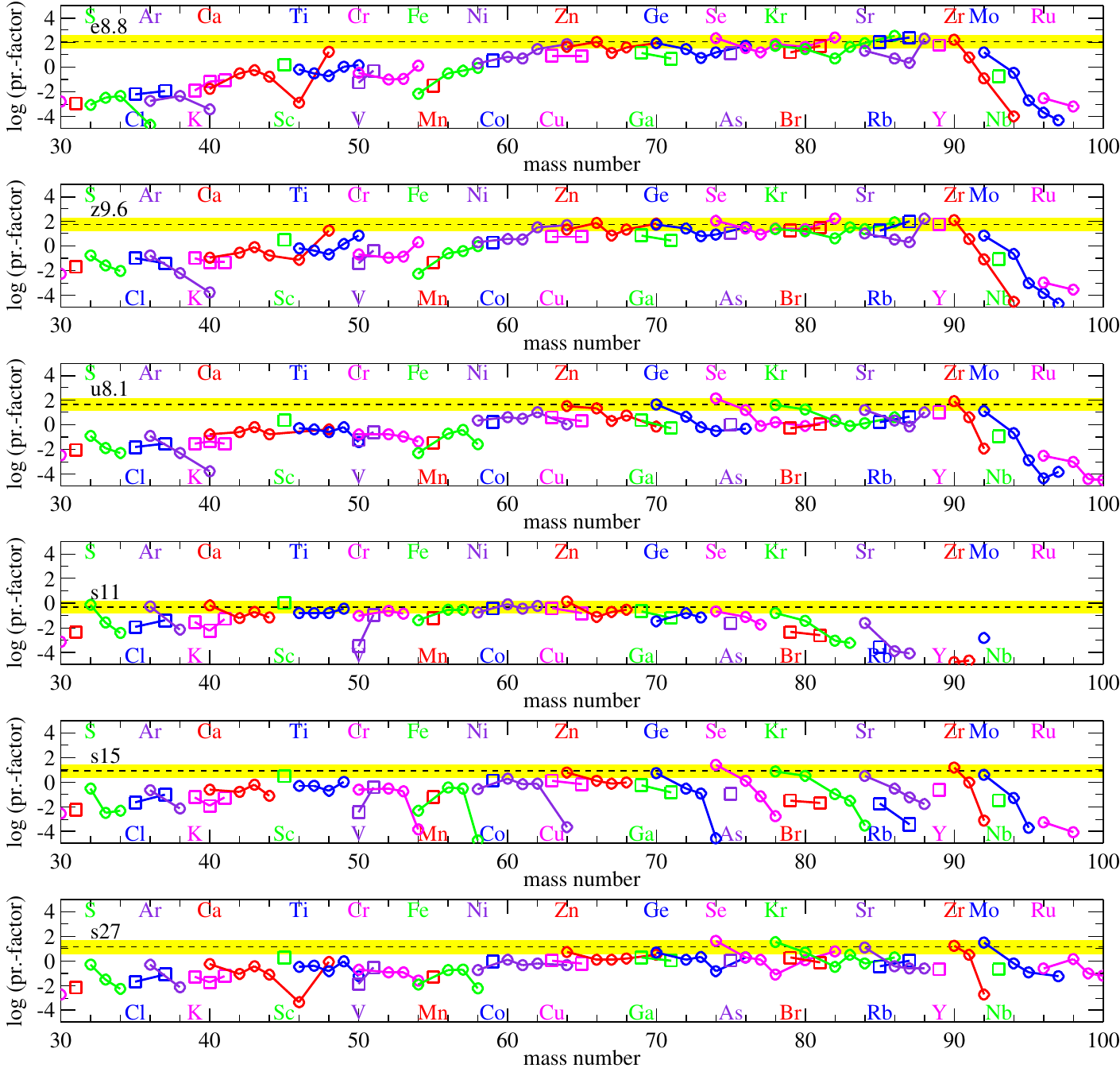}

\caption{Same as Figure~\ref{fig:pfel}, but for isotopic mass
fractions.}

\label{fig:pfiso}
\end{figure*}

The nearly flat
production factors over wide range of $A =64$--90 (except for Ga and As; Fig.~\ref{fig:pfiso}) in e8.8 and z9.6
also suggest that such low-mass SNe are the dominant sources of light
trans-iron elements from Zn to Zr as suggested by
\citet[][]{Wanajo2011a} 
if their mass window is $\Delta M_\mathrm{prog} \sim 0.5$--1. 
This is a consequence of the fact that the
abundant neutron-rich ejecta for these models (Fig.~\ref{fig:yehist})
cover 
the range of $Y_\mathrm{e}$ (down to $\sim 0.40$)
where  the productions of these species become maximal
(Figs.~\ref{fig:yeabun1} and \ref{fig:yeabun2}). By contrast, the
results for model u8.1 (with a similar pre-SN core structure to that of
z9.6) imply that such SNe can only be a source of the proton-rich
isotopes of these trans-iron elements. This is due to the small amount
of neutron-rich ejecta with $Y_\mathrm{e} \sim 0.40$--0.45.

In previous studies, the production of such light trans-iron species has
often been attributed to the weak $s$-process
\citep[e.g.][]{Kaeppeler2011}. \citet{Woosley2007} showed that the
$s$-process in massive stars produces appreciable amounts of light
trans-iron isotopes but only between $A = 65$--85 with a descending trend
of production factors. \X{It also has been claimed that significant
production of these elements can occur in stars above $25\, M_\odot$
via the weak s-process \citep{Pignatari2010}.} \citet{Sukhbold2016} pointed out\X{, however,} that
such light trans-iron species \X{(in particular those above As)} were sizably underproduced in their work  because a
large number of stars formed black holes instead of exploding
and therefore did not contribute to weak $s$-process nuclides. 
\X{A major contribution of the weak s-process
from stars above $25\, M_\odot$ is also unlikely considering
observational evidence against successful explosions of stars
in this mass range \citep{Smartt2015}.}
\X{It should be noted that \citet{Chieffi2013} find that sufficient rotation-induced mixing in massive stars appreciably increases the amount of weak $s$-process elements.}
We speculate that the weak $s$-process in massive stars might still be
responsible for \X{several trans-iron elements, in particular such as} Ga and As, which are underproduced in our models
e8.8 and z9.6 (Figs.~\ref{fig:pfel} and \ref{fig:pfiso}).


\subsection{$^{92}$Mo}\label{subsec:mo92}

\begin{figure*}
\epsscale{1.0}
\plotone{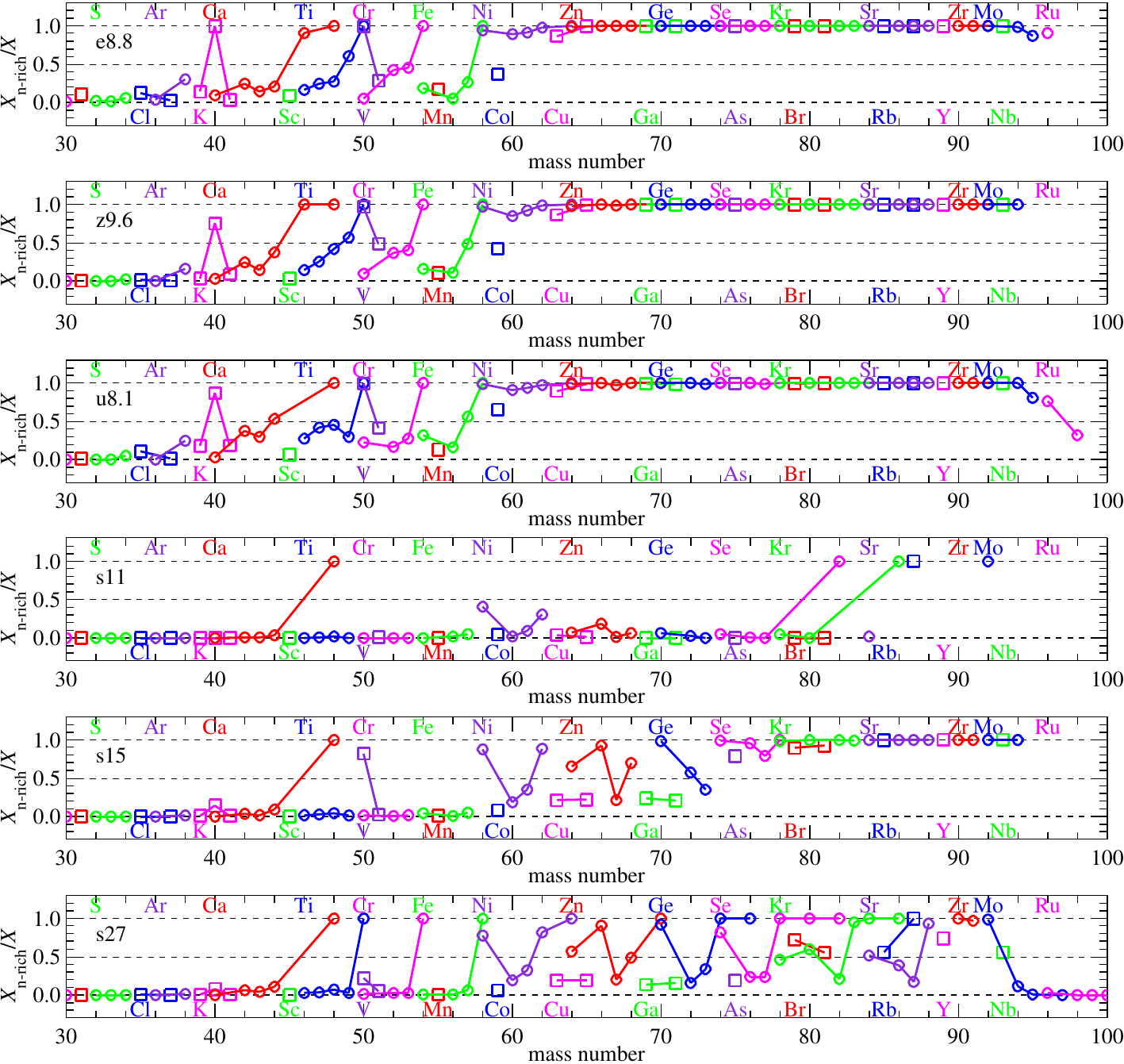}

\caption{Fraction of isotopic yields
    contributed by neutron-rich matter ($Y_\mathrm{e} < 0.4975$,
$M_\mathrm{ej, n}$ in Table~\ref{tab:properties}).
    $X_\mathrm{n-rich}/X = 1.0$ and 0.0 mean that a given
isotope originates exclusively from neutron-rich and proton-rich
ejecta, respectively.   }

\label{fig:npiso}
\end{figure*}

It is interesting to note that model s27 gives abundant $^{92}$Mo
(with the third largest production factor, 30.7, in
Table~3), an important $p$-nuclide whose astrophysical
origin has been a long-lasting problem. In our case $^{92}$Mo is made
in slightly neutron-rich ejecta ($Y_\mathrm{e} \sim 0.47$, see middle
panel of Fig.~\ref{fig:freeze} and middle-bottom panel of
Fig.~\ref{fig:yeabun2}) during nuclear equilibrium \citep[as
suggested by][]{Hoffman1996, Wanajo2006}. With a production factor of
$\sim 30$, such a type of CCSNe, i.e., relatively early
explosions with dense outer envelopes, s27-like explosions
would need to account for $\sim 30\%$ of
all CCSNe to explain the solar abundance of $^{92}$Mo.  A potential problem is that the other $p$-isotope
$^{94}$Mo cannot be made in neutron-rich nuclear equilibrium
\citep{Hoffman1996, Wanajo2006}. The $\nu p$-process in
the neutrino-driven wind may, however, add these isotopes with a high
$^{94}$Mo/$^{92}$Mo ratio \citep{Pruet2006, Wanajo2006, Wanajo2011b}.
Note that we do not find large production factors for the
$p$-isotopes of Ru and Pd as in \citet{Pruet2006}. This is due to the
fact that these isotopes were made by the $\nu p$-process
(\S~\ref{subsubsec:ccap}) in their late-time ($\gtrsim 1$~s) wind ejecta
with high entropies ($S \sim 70\,k_\mathrm{B}$), while
the hydrodynamical simulations of models s15 and s27 stopped at $\sim 0.8$~s after core bounce with entropies still below $S = 50\,k_\mathrm{B}$.



\subsection{$^{56}$Ni}\label{subsec:ni56}

The masses of $^{56}$Ni ejected from all models are listed in
Table~\ref{tab:radio} (7th column), along with those of $^{57}$Ni
(last column). These values, ranging from $\sim 0.002$--$0.006\, M_\odot$,
are about one order of magnitude smaller than typical observed values 
\citep[e.g., $0.07\, M_\odot$ for SN~1987A][]{Bouchet1991}.
The
$^{56}$Ni masses in our study should be taken, however, as lower limits for
the massive models s11, s15, and s27, for which we omitted the outer envelopes
including
large parts of the Si/O layer.  
We therefore likely underestimate the amount of
  $^{56}$Ni produced by explosive nucleosynthesis in the supernova
  shock. Moreover, 2D models face a generic
  difficulty in determining the amount of $^{56}$Ni
  made by explosive burning in the shock, since the
  shocked material is funneled around the neutrino-driven
  outflows onto the proto-neutron star with little mixing by the Kelvin-Helmholtz
  instability into the neutrino-driven ejecta \citep{Mueller2015}.
  The production of $^{56}$Ni by explosive burning also depends
  on the post-shock temperatures, and could be related
  to the slow rise of the explosion
  energy to only (0.3--$1.5) \times 10^{50}$~erg at the end of the simulations.

In addition, the core-collapse simulations of s15
and s27 stopped
at a time when accretion was still ongoing and the
  mass ejection rate in the neutrino-driven outflows
  was still high, so that we may miss some late-time contribution
to $^{56}$Ni in the neutrino-driven ejecta \cite[see, e.g.,][]{Wongwathanarat2016}. 
In fact, the late-time ejecta of s15 and s27 are mostly
proton-rich and are dominated by $^{56}$Ni and $\alpha$-particles
(Fig.~\ref{fig:toseed} and the right-bottom panel of
Fig.~\ref{fig:yeabun2}). 

For the low-mass models e8.8, z9.6, and u8.1, the
listed values can be regarded as the final ones; the ${}^{56}$Ni mass is definitely very small
  ($\sim 0.002$--$0.003\, M_\odot$) for these cases.
It has already been argued before for e8.8 \citep{Kitaura2006,Wanajo2011a,Wanajo2013},
that the small ${}^{56}$Ni may be consistent 
with the value estimated for some  observed low-luminosity supernovae
\citep[e.g.,][]{Hendry2005, Pastorello2007} with small
ejecta masses. The ${}^{56}$Ni mass and the other
explosion properties of the three low-mass models
are also consistent with the remnant composition and reconstructed
light curve of the Crab supernova SN~1054
\citep{Smith2013,Tominaga2013,Moriya2014}. From the nucleosynthetic
point of view, low-mass iron core supernovae thus appear an equally viable
explanation for these events compared to ECSNe.

\subsection{$^{60}$Fe and other radioactive isotopes}\label{subsec:fe60}

The low-mass models e8.8 and z9.6 produce appreciable amounts of $^{60}$Fe,
$3.61 \times 10^{-5}\, M_\odot$ and $3.14 \times 10^{-5}\, M_\odot$,
respectively (6th column in Table~\ref{tab:radio}), which is comparable
to the IMF-averaged ejection mass from CCSNe, (2.70--$3.20) \times
10^{-5}\, M_\odot$ in \citet{Sukhbold2016}. Note that, in massive stars,
$^{60}$Fe is produced by successive neutron captures from iron isotopes
\X{\citep{Timmes1995,Limongi2006}} in the outer envelope, which is not included
in our analysis of the
massive models s11, s15, and s27. In e8.8 and z9.6 $^{60}$Fe forms in
neutron-rich NSE ($Y_\mathrm{e} \sim 0.42$--0.43, right-bottom panel of
Fig.~\ref{fig:yeabun2}) as suggested by \citet{Wanajo2013b}. This
indicates that ECSNe or low-mass CCSNe account for at least $\sim 10\%$
(see $f_\mathrm{e8.8}$ and $f_\mathrm{z9.6}$ in \S~\ref{subsec:isolar})
of live $^{60}$Fe in the Galaxy. Models other than e8.8 and z9.6 produce
little $^{60}$Fe in their innermost ejecta because of their small
masses of neutron-rich ejecta.

Table~\ref{tab:radio} also lists the masses of other important
radioactive isotopes $^{26}$Al, $^{41}$Ca, $^{44}$Ti, and $^{53}$Mn,
which are, however, negligibly small compared to the contribution
  from the outer
envelopes  of massive stars \citep[and from neutrino-driven wind for $^{44}$Ti,][]{Wongwathanarat2016}. Recent work by \citet{Sukhbold2016}
suggested the IMF-averaged amount of $^{26}$Al from massive stars to be
(2.80--$3.63) \times 10^{-5}\, M_\odot$. Their resultant mass ratio of
$^{60}$Fe to $^{26}$Al, $\sim 1$ \citep[although a factor of 2 smaller
than previous estimates, e.g.,][]{Woosley2007}, conflicts with the \R{mass ratio $\sim 0.34$}
inferred from gamma-ray observation \citep[\R{flux ratio $0.148\pm 0.06$},][]{Wang2007}. Taking
the observational value as a constraint, this would
  suggest that  \citet{Sukhbold2016}
overestimated the $^{60}$Fe mass about a factor of 3. The reason for this can be
attributed to uncertainties in the relevant reaction rates
\citep{Woosley2007, Tur2010}, while for our cases $^{60}$Fe forms in
nuclear equilibrium and thus individual reactions are irrelevant. If
this is the case, ECSNe or low-mass SNe can contribute to live $^{60}$Fe
up to $\sim 30\%$ of the amount in the Galaxy.

\begin{figure}
\epsscale{1.0}
\plotone{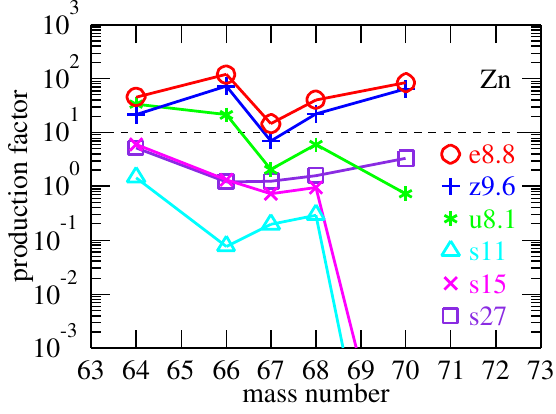}

\caption{Production factors of Zn isotopes (nucleosynthetic yields
relative to solar values) for all models indicated by different
symbols in the legend.}

\label{fig:pfzn}
\end{figure}

\section{Conclusions}\label{sec:conclusions}

We have examined the nucleosynthesis in the innermost ejecta (0.01--$0.03\,
M_\odot$) of CCSNe, including ECSNe, by adopting thermodynamic
trajectories obtained from self-consistent (general-relativistic with one
exception) 2D core-collapse
supernova explosion models with multi-group neutrino transport. We explored the six models e8.8 ($8.8\, M_\odot$ ECSN, $1\,
Z_\odot$), z9.6 ($9.6\, M_\odot$, $0\, Z_\odot$), u8.1 ($8.1\, M_\odot$,
$10^{-4}\, Z_\odot$), s11 ($11.2\, M_\odot$, $1\, Z_\odot$), s15
($15.0\, M_\odot$, $1\, Z_\odot$), and s27 ($27.0\, M_\odot$, $1\,
Z_\odot$), with a progressively shallower core-density gradient in that
order \citep[Fig.~8 in][]{Janka2012}. In this paper, we focused on the
effects of these
differences in the pre-SN core-density structure on the
nucleosynthesis, and did not attempt to address the exact dependence of the
nucleosynthesis on 
mass or metallicity (owing to the limited number of available
models). Our results indicate, however, that ``low-mass'' progenitors
close to the iron core formation limit (which are characterized
by off-center ignition of oxygen burning) stand apart from ``massive stars'' due
to shared structural and evolutionary features: 
The low-mass models e8.8, z9.6, and u8.1 (near the low-mass end of
SN progenitors) have abundant neutron-rich matter (40--50\%) in their
ejecta because of fast shock expansion after the onset of the explosion. By contrast, proton-rich matter
dominates the ejecta of the more massive models s11, s15, and s27 by far, because 
slower shock expansion allows neutrino interactions to efficiently raise
$Y_\mathrm{e}$.

For this reason,  our nucleosynthesis calculations in post-processing steps resulted in
remarkably different outcomes between the low-mass (e8.8, z9.6, and u8.1)
and massive (s11, s15, and s27) models. We determined that  low-mass SNe, in particular
those represented by e8.8 (ECSN) and z9.6, could be the 
dominant source of light trans-iron elements from Zn to Zr (except
for Ga and As), if these events account for $\sim 8$--14\% (or a
mass window of $\Delta M_\mathrm{prog} = 0.5$--$1\, M_\odot$) of all
CCSNe (including ECSNe). These species are made predominantly in
neutron-rich ejecta during nuclear-equilibrium phases. Good agreement
of the nucleosynthetic yields with the solar-abundance pattern over a
wide range of isotopes with $A = 64$--90 strongly supports  our
conclusion, which sets this candidate site apart from other  previously suggested
astrophysical sites, e.g., hypernovae \citep[as origin of
Zn,][]{Umeda2002, Tominaga2007, Nomoto2013} and the weak $s$-process
in massive stars \citep[as origin of light trans-iron
elements,][]{Woosley2007, Kaeppeler2011}. These SNe (e8.8 and z9.6)
could also be important  sources of the neutron-rich isotope $^{48}$Ca and the live
radioactive species $^{60}$Fe in the Galaxy, and supplement the chemogalactic
contribution of other sites such as
rare SNe~Ia \citep{Woosley1997} with high ignition densities  and massive stars
\citep[e.g.,][]{Sukhbold2016}. Model u8.1, however, was found
to have contributions of proton-rich isotopes of light
trans-iron elements only, despite its similar (but slightly 
shallower) core-density gradient compared to that of z9.6.

We found that the innermost ejecta of massive SNe make little
contributions to the chemical inventory of  the Galaxy, except for proton-rich isotopes of light
trans-iron elements (in s15 and s27). The most massive model s27
exhibited an interesting production of $p$-nucleus $^{92}$Mo, which could explain
the solar amount of $^{92}$Mo if such s27-like events 
accounted for $\sim 30\%$ of all CCSNe. This was traced back to a sizable amount of  slightly neutron-rich ejecta ($Y_\mathrm{e} \sim 0.47$) with
moderately high entropies ($\sim 30\, k_\mathrm{B}/\mathrm{nuc}$), in which
$^{92}$Mo is made in nuclear equilibrium \citep{Hoffman1996,
Wanajo2006}. Note that our calculations did not include the outer
envelopes, from which iron-group and lighter elements would be ejected
in these massive SNe. Instead, we focused mostly on the neutrino-processed
  ejecta during the first few hundreds of milliseconds after shock revival.
  This implies that our results on the
  production of iron-group and trans-iron
  elements in the more massive models (s11, s15, s27) are only
  indicative of the overall nucleosynthesis and may miss
  important contributions. For the ECSN progenitor e8.8 and the
  ECSN-like models z9.6 and u8.1, the nucleosynthesis in
  this range is essentially complete, barring minor contributions
from the neutrino-driven wind that follows after our simulations
were terminated.

Aside from the fact that the nucleosynthesis in s11, s15, and s27
is not yet complete, some further caveats should be added to our
conclusions here. Although nucleosynthesis
calculations based on self-consistent 2D core-collapse
supernova models represent a fundamental improvement
over previous 1D studies (which mostly excluded neutrino-processed
ejecta), the explosion dynamics in 3D
has emerged as noticeably different from 2D \citep[see][for recent reviews]{Janka2016,Mueller2016}. The interaction between buoyant, neutrino-heated
ejecta and accretion downflows tends to brake
both outflows and downflows \citep{Melson2015,Mueller2015}
and could therefore modify the final $S$ and $Y_\mathrm{e}$ in the
neutrino-driven ejecta. So far,
model z9.6 offers the only opportunity for a  direct comparison of the nucleosynthesis conditions. For this progenitor, 3D effects
only moderately affect the explosion dynamics
\citep{Melson2015} and may only slightly
raise the minimum $Y_\mathrm{e}$ in the ejecta without fundamentally changing
the neutron-rich nucleosynthesis in this progenitor
\citep{Mueller2016}. The situation for
other low-mass iron core-collapse supernovae and ECSNe is likely similar. 

For more massive progenitors, larger systematic effects
on the nucleosynthesis conditions in 3D cannot
be excluded. The results of \citet{Mueller2015} for
an $11.2 M_\odot$ model suggest that
the outflow velocities (which affect the $Y_\mathrm{e}$) and the terminal
entropies in 3D can be considerably lower, but these results
have yet to be borne out for a broader range of progenitors. The mixing
of shocked material into the neutrino-heated ejecta is likely more
efficient in 3D, and will affect the contribution of explosive
nucleosynthesis in the shock to the total ejecta, because this
matter is swept outward instead of being accreted onto the new-born
neutron star. Along with the
low explosion energies of the massive 2D models, this may explain
the unusually small mass of ${}^{56}$Ni found in our calculations. 

Effects that may change the final $Y_\mathrm{e}$ in the neutrino-processed
ejecta by altering the neutrino emission also warrant further
exploration. Aside from uncertainties in the neutrino microphysics,
the impact of the genuinely three-dimensional LESA (lepton-number emission self-sustained asymmetry)
instability \citep{Tamborra2014, Janka2016} on the nucleosynthesis conditions
deserves investigation. As LESA produces a global
asymmetry in the flux difference between electron neutrinos
and antineutrinos, it will contribute to a spread in the $Y_\mathrm{e}$-distribution
in the ejecta. A detailed analysis of LESA effect
in z9.6 is currently underway; the neutron-rich bubbles in the
early ejecta
remain a robust feature for this progenitor also in the presence
of LESA (Melson et al., in preparation).

In summary, while our conclusions for low-mass SNe might be robust,
those for massive models should only  be taken as suggestive.
Eventually, 
successful explosion models of such massive SNe in 3D are required
for drawing firmer conclusions.
It is, however, also important to scan the low-mass end
of the progenitor range with a finer resolution in progenitor mass. 
While
\citet{Wanajo2011a, Wanajo2013a, Wanajo2013b} suggested that ECSNe
could be important contributors of light trans-iron elements
(including Zn), $^{48}$Ca, and $^{60}$Fe, our results show that CCSNe
near the low-mass end with iron-cores (z9.6) have almost the same
nucleosynthetic outcomes. A model (u8.1) with only a slightly
shallower core-density gradient resulted in, however, a substantially
different result. Stellar evolution models and core-collapse
supernova simulations therefore need to better address over which
mass range the progenitors possess core-density profiles that
are sufficiently steep
to produce ECSN-like nucleosynthesis near the low-mass end
of the progenitor-mass spectrum.

It should also be noted that we neglect late-time ejecta
in the neutrino-driven wind in the present study.
These could also  contribute to the
enrichment of trans-iron species to the Galaxy by a weak $r$-process
\citep[e.g.,][]{Wanajo2013} and a $\nu p$-process
\citep{Froehlich2006, Pruet2006, Wanajo2006}. Finally, our result
should be tested by a study of Galactic chemical evolution to
reproduce the observational signatures of light trans-iron elements in
the Galaxy.

\acknowledgements

\X{S.W. thanks N.\ Prantzos and M.\ Limongi for discussions about
the weak s-process in massive stars.}
This work was supported by the RIKEN iTHES Project (SW), by the JSPS
Grants-in-Aid for Scientific Research (26400232, 26400237; SW),
\X{by JSPS and CNRS under the Japan-France Research Cooperative Program (SW),}
by the Deutsche Forschungsgemeinschaft through the
Excellence Cluster Universe EXC 153 (TJ), and by the European Research
Council through grant ERC-AdG No.~341157-COCO2CASA (TJ).
We also acknowledge partial support by the
Australian Research Council through a Discovery Early Career
Researcher Award DE150101145 (BM) and ARC Future Fellowships FT120100363 (AH) and FT160100035 (BM)
and partial support by the US National Science Foundation under Grant No.\ PHY-1430152 
(JINA Center for the Evolution of the Elements).  This
research was undertaken with the assistance of resources from
the Max Planck Computing and Data Facility (using the
\emph{hydra} cluster), the Minnesota Supercomputing Institute, the
National Computational Infrastructure (NCI), which is supported by the
Australian Government, and the
Pawsey Supercomputing Centre with funding from the Australian
Government and the Government of Western Australia.

\end{document}